\newcommand*\diff{\mathop{}\!\mathrm{d}}

\documentclass[%
 reprint,
nofootinbib,
 amsmath,amssymb,
 aps,
]{revtex4-2}
\usepackage{tensor}
\usepackage{graphicx}
\usepackage{dcolumn}
\usepackage{bm}
\usepackage{comment}
\usepackage[table]{xcolor}
\usepackage{booktabs}

\usepackage{braket}
\newcommand{\virgolette}[1]{``#1''}
\usepackage{diagbox}

\begin{document}

\preprint{APS/123-QED}

\title{Quantum fluctuations in the effective relational GFT cosmology} 

\author{L. Marchetti}
\email[]{luca.marchetti@phd.unipi.it, luca.marchetti@pi.infn.it}
\affiliation{Università di Pisa,\\Lungarno Antonio Pacinotti 43, 56126 Pisa, Italy, EU}
\affiliation{Ludwig-Maximilians-Universit\"at München,\\ Ludwig-Maximilians-Universit\"at München \\ Theresienstrasse 37, 80333 M\"unchen, Germany, EU}
\affiliation{Istituto Nazionale di Fisica Nucleare sez. Pisa,\\Largo Bruno Pontecorvo 3, 56127 Pisa, Italy, EU}

\author{D. Oriti}
\email[]{daniele.oriti@physik.lmu.de}
\affiliation{Arnold Sommerfeld Center for Theoretical Physics, \\ Ludwig-Maximilians-Universit\"at München \\ Theresienstrasse 37, 80333 M\"unchen, Germany, EU}

\date{\today}

\begin{abstract}
We analyze the size and evolution of quantum fluctuations of cosmologically relevant geometric observables, in the context of the effective relational cosmological dynamics of GFT models of quantum gravity. We consider the fluctuations of the matter clock observables, to test the validity of the relational evolution picture itself. Next, we compute quantum fluctuations of the universe volume and of other operators characterizing its evolution (number operator for the fundamental GFT quanta, effective Hamiltonian and scalar field momentum). In particular, we focus on the late (clock) time regime, where the dynamics is compatible with a flat FRW universe, and on the very early phase near the quantum bounce produced by the fundamental quantum gravity dynamics.
\end{abstract}

\maketitle
\section{Introduction}
Three closely related challenges have to be overcome by fundamental quantum gravity approaches, especially those based on discrete or otherwise non-geometric, non-spatiotemporal entities, in order to make contact with General Relativity and observed gravitational physics, based on effective (quantum) field theory. The first is the continuum limit/approximation leading from the fundanental entities and their quantum dynamics to an effective continuum description of spacetime and geometry, with matter fields living on it \cite{Oriti:2017ave}. This requires a mixture of renormalization analysis of the fundamental quantum dynamics and of coarse-graining of its states and observables. The second is a classical limit/approximation of the sector of the theory corresponding to (would-be) spacetime and geometry, to show that indeed an effective classical dynamics compatible with General Relativity and observations emerges, once in the continuum description \cite{Oriti:2017ave}. The third is a definition of suitable observables that can, on the one hand, give a physical meaning to both continuum and classical approximations in terms of spacetime geometry and gravity, and, on the other hand, allow to make contact with phenomenology \cite{Marchetti:2020umh}. In particular, suitable observables are needed to recast the dynamics of the quantum gravity system, in the same continuum and classical approximations, at least, in more customary local evolutionary terms, i.e. in the form of evolution of local quantities with respect to some notion of time \cite{Tambornino:2011vg, Giddings:2005id}. This, in fact, is the language of effective field theory used in gravitational and high energy physics. The first two challenges are standard in any quantum many-body system, but are made more difficult in the quantum gravity context by the necessary background (and spacetime) independence of the fundamental theory, which requires adapting non-trivially standard renormalization, coarse-graining and classical approximation techniques. The same background independence, closely related at the formal level to the diffeomorphism invariance of General Relativity \cite{Giulini:2006yg}, makes the third challenge a peculiar difficulty in quantum gravity. Locality or temporal evolution cannot be defined with respect to any manifold point or direction and generic configurations (classical and even less quantum) of the gravitational field (or what replaces it at the fundamental level) do not single out any such notions either. Beside special situations (e.g. in the presence of asymptotic boundaries) local and temporal geometric observables can be understood as relational quantities, i.e. defined as a relation between geometry and other dynamical matter degrees of freedom, that provide a notion of local regions and temporal direction when used as physical reference frames. In other words, and restricting to the issue of time \cite{Kuchar2011, Isham1992}, the relational perspective holds that the absence of preferred, external or background notions of time in generally relativistic quantum theories does not mean that there is no quantum evolution, but only that evolution should be defined with respect to internal, physical degrees of freedom \cite{Hoehn:2019owq,Tambornino:2011vg}.

From the perspective of \virgolette{Quantum General Relativity} theories \cite{Rovelli:2004tv,Thiemann:2007pyv}, in which the fundamental entities remain (quantized) continuum fields, the relational strategy to define evolution boils down to either the selection of a relational clock at the classical level, in terms of which the remaining subsystem is canonically quantized (\virgolette{tempus ante quantum} \cite{Isham1992,Anderson:2010xm}) or the definition of an appropriate clock-neutral quantization (e.g., Dirac quantization) and the representation of classical complete (i.e., relational) observables \cite{Tambornino:2011vg, Dittrich:2004cb, Dittrich:2005kc, Rovelli:2001bq} on the physical Hilbert space resulting from such quantization (\virgolette{tempus post quantum} \cite{Isham1992,Anderson:2010xm}). Of course, while the first approach (deparametrization) is technically easier, when possible, the second one is in principle preferable because manifestly \virgolette{clock covariant}, since it treats all the quantum degrees of freedom on the same footing, thus allowing in principle to switch from one relational clock to another (see \cite{Bojowald:2010xp, Bojowald:2010qw, Hoehn2011, Hoehn2018} for more details).

In \virgolette{emergent quantum gravity} theories, in which the fundamental degrees of freedom are pre-geometric and non-spatiotemporal, and not identified with (quantized) continuum fields, the situation has an additional layer of complications \cite{Marchetti:2020umh}. Any kind of continuum notion in such theories is expected to emerge in a {\it proto-geometric phase} of the theory from the collective behavior of the fundamental entities, i.e. only at an effective and approximate level. Among such continuum notions there is any notion of relational dynamics, as we understand it from the generally relativistic perspective. 

\

In the tensorial group field theory formalism (TGFT) for quantum gravity (see \cite{Oriti:2011jm, Krajewski:2012aw, Carrozza:2016vsq, Gielen:2016dss} for general introductions), comprising random tensor models, tensorial field theories and group field theories (closely related to canonical loop quantum gravity, and providing a reformulation of lattice gravity path integrals and spin foam models), we are in this last emergent spacetime situation \cite{Oriti:2018dsg}. 

The issue of the continuum limit is tackled adapting standard renormalization group  \cite{Carrozza:2016vsq} and statistical methods for quantum field theories, leading also to several results concerning the critical behaviour of a variety of models.
In the more quantum geometric group field theory (GFT) models \cite{Carrozza:2017vkz,Geloun:2016qyb,Magnen:2009at,Baratin:2013rja,Carrozza:2016tih,BenGeloun:2018ekd} (see also \cite{Finocchiaro:2020fhl} and references therein), one can take advantage of the group theoretic data and of their discrete geometric interpretation to give tentative physical meaning to suitable quantum states and to specific regimes of approximation of their quantum dynamics \cite{Oriti:2015qva}. Specifically, the hydrodynamic regime of models of 4d quantum geometry admits a cosmological interpretation and has been analyzed in some detail for simple condensate states \cite{Gielen:2016dss, Gielen:2013naa, Oriti:2016qtz}. The corresponding effective dynamics has been recast in terms of cosmological observables both via the relational strategy and by a deparametrization with respect to the added matter degrees of freedom \cite{Marchetti:2020umh, Oriti:2016qtz, Wilson-Ewing:2018mrp,Gielen:2016dss, Pithis:2019tvp,Oriti:2016acw}. Among many results \cite{Gielen:2014ila,Adjei:2017bfm,Gielen:2019kae, Gielen:2020abd,Pithis:2016wzf,Pithis:2016cxg,Gielen:2020abd}, the correct classical limit in terms of a flat FRW universe has been obtained rather generically for large expectation values of the volume operator at late relational (clock) times \cite{Marchetti:2020umh,Oriti:2016qtz,Gielen:2019kae}, and the big bang singularity is resolved, with a similar degree of generality \cite{Oriti:2016qtz,Gielen:2019kae}, and replaced with a quantum bounce. In addition, the fundamental quantum gravity interactions seem to be able produce (at least for some regime of parameters) an accelerated cosmological expansion, possibly long-lasting, without introducing additional matter (e.g. inflaton) fields \cite{deCesare:2016rsf}.  

The above results have been obtained looking at the expectation values of interesting cosmological observables in (the simplest) GFT condensate states. A careful analysis of quantum fluctuations of the same observables is then important to test the validity of the hydrodynamic description in terms of expectation values, in particular in the large volume limit when one expects classical GR to be valid, but also close to the big bounce regime where one expects them to be strong but still controllable if the bouncing scenario is to be trustable at all. Moreover, the relational evolution relies on the chosen physical (matter) degrees of freedom to behave nicely enough to serve as a good clock, and this would not be the case if subject to strong quantum fluctuations. This analysis of quantum fluctuations is what we perform in this paper. 

The precise context in which we perform the analysis is that of the effective relational dynamics framework developed in \cite{Marchetti:2020umh}. 

This construction is motivated by the argued usefulness and conceptual importance of effective approaches to relational dynamics \cite{Bojowald:2009zzc,Bojowald:2009jj,Bojowald:2010qw,Bojowald:2010xp,Bojowald:2012xy}, and it was suggested a general framework in which the latter is realized in a \virgolette{tempus post quantum} approach, but only at a proto-geometric level, i.e. after some suitable coarse graining, the one provided by the GFT hydrodynamic approximation (or its improvements).

Besides its conceptual motivations, this effective relational framework improves on previous relational constructions in GFT cosmology providing a mathematically more solid definition of relational observables, allowing the explicit computation of quantum fluctuations, which will be one the main objectives of the present work.

This improved effective relational dynamics was obtained by the use of \virgolette{Coherent Peaked States} (CPSs), in which the fundamental GFT quanta collectively (and only effectively) reproduce the classical notion of a spacelike slice of a spacetime foliation labelled by a massless scalar field clock. For this effective foliation to be meaningful quantum flctuations of the clock observables should be small enough (e.g. in the sense of relative variances). When this is the case, the relevant physics is captured by averaged relational dynamics equations for the other observables of cosmological interest, like the universe volume or the matter energy density or the effective Hamiltonian. The purpose of this paper is to explore under which conditions this averaged relational dynamics is meaningful and captures the relevant physics, checking quantum fluctuation for both clock observables and cosmological, geometric ones. 
\section{Effective relational framework for GFT condensate cosmology}\label{sec:erdframework}

The GFT condensate cosmology framework is based on three main ingredients (see e.g.\ \cite{Gielen:2016dss} for a review):
\begin{enumerate}
    \item the identification of appropriate states which admit an interpretation in terms of (homogeneous and isotropic) cosmological $3$-geometries;
    \item the construction of an appropriate relational framework allowing to describe e.g.\ the (averaged) geometric quantities (in the homogeneous and isotropic case, the volume operator) as a function of a matter field (usually a minimally coupled massless scalar field);
    \item the extraction of a mean field dynamics from the quantum equations of motion of the microscopic GFT theory, which in turn determines the relational evolution of the aforementioned (averaged) volume operator.
\end{enumerate}
In this section, we will review the concrete realization of the first two steps and of the first part of the third step (i.e.\ the extraction of a mean field dynamics),
in order to prepare the ground for the calculation of expectation values, first, and the  quantum fluctuations of geometric observables of cosmological interest. More precisely, the first step will be reviewed in Subsection \ref{subsec:reviewgft}, while the second and the first part of the third one will be discussed in Subsections  \ref{subsec:reviewcps}, and \ref{subsec:redwavefunctiondynamics}, respectively. The second part of the last step, which requires the detailed computations of expectation values performed in Subsection \ref{subsec:explicitexpect}, will be instead discussed in Subsection \ref{subsec:effectivevolumecps}.
\subsection{The kinematic structure of GFT condensate cosmology}\label{subsec:reviewgft}
In the Group Field Theory (GFT) formalism \cite{Oriti:2011jm, Krajewski:2012aw, Carrozza:2016vsq, Gielen:2016dss}, one aims at a microscopic description of spacetime in terms of simplicial building blocks \cite{Reisenberger:2000zc}. The behavior of the fundamental \virgolette{atoms} that spacetime has dissolved into is described by a (in general, complex) field $\varphi: G^d\to\mathbb{C}$ defined on $d$ copies of a group manifold, $\varphi(g_{I})\equiv \varphi(g_1,\dots, g_d)$. By appropriate choices of the dimension $d$, of the group manifold $G$, of the combinatorial pairing of field arguments in the action, and of course its functional form, the perturbative expansion of the theory produces amplitudes that can be seen as a simplicial gravity path-integral \cite{Baratin:2011hp}, with the group-theoretic data entering as holonomies of a discrete gravitational connection. Concretely, most 4d gravity models use $d=4$ (i.e., the spacetime dimension), and $G=\text{SL}(2,\mathbb{C})$ (local gauge group of gravity), its Euclidean version, $\text{Spin}(4)$, or $\text{SU}(2)$, once an appropriate embedding into $\text{SL}(2,\mathbb{C})$ or $\text{Spin}(4)$ is specified. This latter choice allows for an explicit connection of the GFT quantum states with those in the kinematical Hilbert space of LQG \cite{Gielen:2016dss}. From now on, therefore, we will specialize to $d=4$ and $G=\text{SU}(2)$. 

Indeed, in this case, the fundamental quanta of the field, assuming it satisfies the \virgolette{closure} condition $\varphi(g_I)=\varphi(g_I h)$ for each $h\in G$, can be interpreted as $3$-simplices (tetrahedra) whose faces are decorated with an equivalence class of geometrical data $[\{g_I\}]=\{\{g_I h\},h\in G\}$ or, in the dual picture, as \emph{open spin-networks}, i.e., nodes from which four links are emanating, each of which is associated to group-theoretical data. From this dual perspective, the closure condition becomes the imposition of invariance under local gauge transformations which act on the spin-network vertex. 
\subsubsection{The GFT Fock space}
The Fock space of such \virgolette{atoms of space} can be constructed in terms of the field operators $\hat{\varphi}(g_I)$ and $\hat{\varphi}^\dagger(g_I)$ subject to the following commutation relations: 
\begin{subequations}
\begin{align}\label{eqn:basiccommutator}
[\hat{\varphi}(g_{I}),\hat{\varphi}^\dagger(g_{I}')]&=\mathbb{I}_{G}(g_{I},g_{I}')\,,\\ [\hat{\varphi}(g_{I}),\hat{\varphi}(g_{I}')]&=[\hat{\varphi}^\dagger(g_{I}),\hat{\varphi}^\dagger(g_{I}')]=0\,,
\end{align}
\end{subequations}
together with a vacuum state $\ket{0}$ annihilated by $\hat{\varphi}$, so that the action of $\hat{\varphi}^\dagger(g_I)$ on $\ket{0}$ creates a \virgolette{quantum of space} with (an equivalence class of) geometric data $\{g_I\}$. The right-hand-side of equation \eqref{eqn:basiccommutator} represents the identity in the space of gauge-invariant (i.e., right diagonal invariant) fields \cite{Gielen:2013naa}.

GFT \virgolette{$(m+n)$-body operators}  $\hat{O}_{n+m}$ 
\begin{align}\label{eqn:secondquantizationoperator}
\hat{O}_{n+m}\equiv \int (\diff g_{I})^m(\diff h_{I})^n\,&O_{m+n}(g_{I}^1,\dots, g_{I}^m,h_{I}^1,\dots, h_{I}^n)\nonumber\\
&\times\prod_{i=1}^m\hat{\varphi}^\dagger(g_{I}^i)\prod_{j=1}^n\hat{\varphi}(h_{I}^j)\,,
\end{align}
are then constructed from the matrix elements $O_{m+n}$, whose form can be determined from simplicial geometric or canonical approaches like Loop Quantum Gravity (LQG) \cite{Ashtekar:2004eh,Rovelli:2004tv, Thiemann:2007pyv}. The same kind of construction can of course be performed in any representation of the relevant Hilbert space. We will work with explicit examples of such operators (number operator, volume operator, massless scalar field operator, etc.~) in the cosmological context.  
\paragraph*{Coupling to a massless scalar field.}
Following \cite{Marchetti:2020umh, Oriti:2016qtz}, a scalar field is minimally coupled to the discrete quantum geometric data, with the purpose of using it as a relational clock at the level of GFT hydrodynamics. This is done by adding to the GFT field and action the degree of freedom associated to a scalar field in such a way that the GFT partition function, once expanded perturbatively around the Fock vacuum, can be identified with the (discrete) path-integral of a model
of simplicial gravity minimally coupled with a free massless scalar field (or, equivalently,
with the corresponding spin-foam model)\footnote{This procedure can in fact be seen as a discrete version of what would be done in a 3rd quantized framework for quantum gravity; indeed, GFT models (like matrix models for 2d gravity) are a discrete realization of the 3rd quantization idea \cite{Gielen:2011dg}.} \cite{Oriti:2016qtz}. Therefore, the field operator changes as follows: 
\begin{equation}
\hat{\varphi}(g_I)\quad\longrightarrow\quad\hat{\varphi}(g_I,\chi)\,,
\end{equation}
meaning that the one-particle Hilbert space is now enlarged to $L^2(\text{SU}(2)^4/\text{SU}(2)\times \mathbb{R})$. So, each GFT atom can carry (in the appropriate basis) a value of the scalar field, which is \virgolette{discretized} on the simplicial structures associated to GFT states and (perturbative) amplitudes \cite{Li:2017uao}. This implies that the commutation relations in \eqref{eqn:basiccommutator} need to be modified consistently, obtaining 
\begin{equation}\label{eqn:frozencommutations}
\left[\hat{\varphi}(g_I,\chi),\hat{\varphi}^\dagger(h_I,\chi')\right]=\mathbb{I}_G(g_I,h_I)\delta(\chi-\chi')\,.
\end{equation} 
and that operators \eqref{eqn:secondquantizationoperator} in the second quantization picture  now involve integrals over the possible values of the massless scalar field \cite{Oriti:2016qtz,Marchetti:2020umh}. 
\subsubsection{GFT condensate cosmology: kinematics}\label{subsub:gftcondensatecosmology}
The Fock space construction described above proves technically very useful in order to address the problem of extraction of continuum physics from GFTs. In particular, in previous works  \cite{Gielen:2014ila, Oriti:2016qtz}, this was exploited to build quantum states that, in appropriate limits, can be interpreted as continuum and homogeneous $3$-geometries, thus paving the way to cosmological applications of GFTs. Such states are characterized by a single collective wavefunction, defined over the space of geometries associated to a single tetrahedron or, equivalently (when some additional symmetry conditions are imposed on the wavefunction) over the minisuperspace of homogeneous geometries \cite{Gielen:2014ila}. For such condensate states then, classical homogeneity is lifted at the quantum level by imposing \lq wavefunction homogeneity\rq. Among the many possible condensate states (characterized by different \virgolette{gluing} of the fundamental GFT quanta one to another) satisfying the above wavefunction homogeneity, most of the attention was directed towards the simplest GFT coherent states, i.e.,
\begin{equation}\label{eqn:coherentstates}
    \ket{\sigma}=N_\sigma\exp\left[\int \diff \chi\int\diff g_I\,\sigma(g_I,\chi)\hat{\varphi}^\dagger(g_I,\chi)\right]\ket{0},
\end{equation}
where
\begin{subequations}
\begin{align}
N_\sigma&\equiv e^{-\Vert \sigma\Vert^2/2},\\\
\Vert\sigma\Vert^2&=\int \diff g_I\diff\chi\vert\sigma(g_I,\chi)\vert^2\,.
\end{align}
\end{subequations}
They satisfy the important property
\begin{equation}\label{eqn:eigenstateannihilation}
    \hat{\varphi}(g_I,\chi)\ket{\sigma}=\sigma(g_I,\chi)\ket{\sigma}\,,
\end{equation}
i.e., they are eigenstates of the annihilation operator.

In order to make contact with cosmological geometries, one typically also imposes isotropy on the wave function, requiring the associated tetrahedra to be equilateral. This results in the following condensate wavefunction \cite{Oriti:2016qtz}
\begin{equation*}
    \sigma(g_I,\chi)=\sum_{j=0}^\infty\sigma_{j,\vec{m},\iota_+}\mathcal{I}^{jjjj,\iota_+}_{n_1n_2n_3n_4}\sqrt{d^4(j)}\prod_{i=1}^4D^j_{m_in_i}(g_i)\,,
\end{equation*}
where
\begin{equation*}\label{eqn:sigmachi}
    \sigma_{\{j,\vec{m},\iota_+\}}(\chi)=\sigma_j(\chi)\overline{\mathcal{I}}^{jjjj,\iota_+}_{m_1m_2m_3m_4}\,.
\end{equation*}
and where $d(j)=2j+1$, $j$ are spin labels, $D^j_{mn}$ are Wigner representation matrices, $\mathcal{I}$ are intertwiners, and $\mathcal{I}^{jjjj,\iota_+}_{m_1m_2m_3m_4}$ is an eigenvector of the LQG volume operator with the largest eigenvalue \cite{Oriti:2016qtz}. After imposition of isotropy, $\sigma_j$ becomes the quantity effectively encoding the physical properties of the state.

\subsection{Effective relational dynamics framework and its implementation in GFT condensate cosmology}\label{subsec:reviewcps}
In \cite{Marchetti:2020umh}, a procedure for extracting an effective relational dynamics framework was proposed for the cosmological context, when one is interested in describing the evolution of some geometric operators with respect of some scalar matter degree of freedom. Since our analysis of quantum fluctuations will take place within such effective relational framework, let us summarize how it is obtained and under which conditions it is expected to be meaningful. In the following, we will analyze also the limits of validity of such conditions. 
\subsubsection{Effective relational evolution of geometric observables with respect to scalar matter degrees of freedom}\label{subsub:generalerd}
The fundamental observables one is interested in are: a \virgolette{scalar field operator} $\hat{\chi}$, a set of \virgolette{geometric observables}\footnote{For instance, in a cosmological context in which one is interested only to homogeneous and isotropic geometries, the volume operator is expected to capture all the geometric properties of the system. In this case, therefore, one only includes this volume operator among the geometric observables of interest.} $\{\hat{O}_a\}_{a\in S}$ and a \virgolette{number operator} $\hat{N}$, counting the number of fundamental \virgolette{quanta of space}. Since one is assuming that the theory, at this pre-geometric level, is entirely clock-neutral (and so are all the operators above), the effective relational dynamics is realized through an appropriate choice of a class of states $\ket{\Psi}$ having both an intepretation in terms of continuum geometries\footnote{In this sense, the operators $\hat{\chi}$ and $\{\hat{O}_a\}_{a\in O}$ are expected to have an interpretation in terms of scalar field and geometric quantities respectively, only when averaged on such states.} (and thus possibly characterizing a proto-geometric phase of the theory) and also carrying a notion of relationality. More precisely, they should allow for the existence of an Hamiltonian operator $\hat{H}$ such that, for each geometric observable $\hat{O}_a$,
\begin{subequations}\label{eqn:averagedrelationaldynamics}
\begin{equation}\label{eqn:averageddynamics}
i\frac{\diff}{\diff\braket{\hat{\chi}}_{\Psi}}\braket{\hat{O}_{a}}_{\Psi}=\braket{[\hat{H},\hat{O}_{a}]}_{\Psi}\,,
\end{equation}
at least locally and far enough from singular turning points of the scalar field clock\footnote{The above equation is however expected to hold globally if the clock is a minimally coupled massless scalar field, which is going to be the only case we will consider here.}.
In order to interpret this evolution as truly relational with respect to the scalar field used as a clock, all the moments of $\hat{H}$ and of the scalar field momentum $\hat{\Pi}$ on $\ket{\Psi}$ should be equal. In particular, this implies that the averages of these two operators on $\ket{\Psi}$ should be equal,
\begin{equation}\label{eqn:piequalh}
\braket{\hat{H}}_{\Psi}=\braket{\hat{\Pi}}_{\Psi}\,.
\end{equation}
\end{subequations}
This equality was investigated in \cite{Marchetti:2020umh} in the context of GFT cosmology, and we will discuss it further below. 

A further condition that is necessary in order to interpret equation \eqref{eqn:averageddynamics} as a truly relational dynamics involves the smallness of the quantum fluctuations on the matter clock. In \cite{Marchetti:2020umh}, this was imposed by requiring the relative variance of $\hat{\chi}$ on $\ket{\Psi}$ to be much smaller than one, and to have the characteristic many-body $\braket{\hat{N}}^{-1}$ behavior, i.e., 
\begin{equation}\label{eqn:smallfluctuationsclock}
    \delta^2_\chi\ll 1\,,\qquad \delta^2_\chi\sim \braket{\hat{N}}^{-1}\,,
\end{equation}
where the relative variance on $\ket{\Psi}$ is defined as
\begin{equation*}
\delta^2_{O}=\frac{\braket{\hat{O}^2}_{\Psi}-\braket{\hat{O}}^2_{\Psi}}{\braket{\hat{O}}^2_{\Psi}}\,.
\end{equation*}
This is of course formally correct only when one is assuming that the expectation value of $\hat{\chi}$ is non-zero, as we will discuss further below. When this is not the case, one should define some thresholds which the relative variances should be smaller than \cite{Marchetti:2020umh}.  

Let us also notice, that, strictly speaking, one would have to require that all the moments of the scalar field operator higher than the first one are much smaller than one in order to guarantee a negligible impact of quantum fluctuations of the clock on the relational framework. However, one also expects that, being the system fundamentally a many-body system (for which the second condition in \eqref{eqn:smallfluctuationsclock} is satisfied), moments higher than the second one get also suppressed in the large $N$ limit which we will be mainly interested in, forming a hierarchy of less and less important quantum effects (typically, in many-body systems, relative moments of order $n$ are suppressed by $\braket{N}^{-(n-1)}$, with $n>1$). In the asymptotic $N\to\infty$ regime, therefore, one should be allowed to characterize quantum fluctuations essentially by the behavior of relative variances. This might not be the case, on the other hand, in intermediate regimes of smaller $N$, where indeed there is no good reason to believe such a hierarchy to be realized. In such cases the impact of quantum fluctuations has to be studied more carefully.
\subsubsection{Implementation in GFT condensate cosmology: CPSs}\label{subsub:cps}
The strategy to realize the above framework in the context of GFT condensate cosmology in \cite{Marchetti:2020umh} made use of Coherent Peaked States (CPSs). These states are constructed so that they can provide, under appropriate approximations, \virgolette{bona fide} leaves of a relational $\chi$-foliation of spacetime. Given the proto-geometric nature of the states \eqref{eqn:coherentstates} the idea is to look for a subclass of them characterized by a given value of the relational clock, say $\chi_0$, so that the GFT quanta collectively conspire to the approximate reconstruction of a relational leaf of spacetime labelled by $\chi_0$ itself. Since, in the condensate states \eqref{eqn:coherentstates} the information about the state is fully encoded in the condensate wavefunction, in \cite{Marchetti:2020umh} relational proto-geometric states are chosen among those where this wavefunction has a strong peaking behavior: 
\begin{equation}\label{eqn:wavefunctioncps}
 \sigma_{\epsilon}(g_I,\chi)\equiv \eta_{\epsilon}(g_I;\chi-\chi_{0},\pi_{0})\tilde{\sigma}(g_I,\chi)\,,
 \end{equation} 
where $\eta_{\epsilon}$ is the so-called \emph{peaking function} around $\chi_{0}$ with a typical width given by $\epsilon$. For instance, one can choose
a Gaussian form
\begin{equation}\label{eqn:peakingfunction}
\eta_{\epsilon}(\chi-\chi_{0},\pi_{0})\equiv \mathcal{N}_{\epsilon}\exp\left[-\frac{(\chi-\chi_{0})^2}{2\epsilon}\right]\exp[i\pi_{0}(\chi-\chi_{0})]\,,
\end{equation}
where $\mathcal{N}_{\epsilon}$ is a normalization constant and where it was assumed that the peaking function does not depend on the group variables $g_{I}$ (the dependence on quantum geometric data is therefore fully encoded in the remaining contribution to the full wavefunction). Further, the \emph{reduced wavefunction} $\tilde{\sigma}$ was assumed not to spoil the peaking properties\footnote{For instance, a reduced wavefunction (whose modulus is) behaving as $\exp[\chi^n]$ with $n\ge 2$ would certainly destroy any localization property of the wavefunction $\sigma_\epsilon$. On the other hand, any function (whose modulus is) characterized by polynomial or exponential $\exp\chi$ behavior would be an admissible candidate for the reduced condensate wavefunction.} of $\eta_\epsilon$ \cite{Marchetti:2020umh}. Since the reduced wavefunction is determined dynamically (see Subsection \ref{subsec:redwavefunctiondynamics} below), this constrains the space of admissible solutions to the dynamical equations. However, in the cosmological case, this will not result on discarding any solution, since the most general one (see equation \eqref{eqn:generalsolution}) has the desired property.

In order for the average clock value to be really meaningful in defining a relational evolution, it is necessary for the width $\epsilon$ of the peaking function to be small, $\epsilon\ll 1$. However, as remarked in \cite{Marchetti:2020umh} and as we will see explicitly below, taking the limit $\epsilon\to 0$, would of course make quantum fluctuations on the momentum of the massless scalar field clock to diverge, thus making the clock highly quantum even in regimes in which we expect to reach some kind of semi-classicality. Moreover, even by considering a small but finite $\epsilon$, there is no guarantee in principle for quantum fluctuations in the scalar field momentum to become controllable in the same semi-classical regime. This can be ensured, however, by imposing the additional condition 
\begin{equation}\label{eqn:epi}
    \epsilon\pi_0^2\gg 1\,.
\end{equation}
For more remarks and comments on this particular class of states we refer to \cite{Marchetti:2020umh}.
\subsection{Reduced wavefunction dynamics and solutions}\label{subsec:redwavefunctiondynamics}
Since the relational approach discussed in the previous section is by its very nature effective and approximate, following \cite{Marchetti:2020umh} we will only extract an effective mean field dynamics from the full quantum equations of motion. In other words, we will only consider the imposition of the quantum equations of motion averaged on the states that we consider to be relevant for an effective relational description of the cosmological system: 
\begin{align}\label{eqn:simplestschwinger}
&\left\langle\frac{\delta S[\hat{\varphi},\hat{\varphi}^\dagger]}{\delta\hat{\varphi}^\dagger(g_I,\chi_{0})}\right\rangle_{\sigma_{\epsilon};\chi_{0},\pi_{0}}\nonumber\\
&\quad\equiv\left\langle\sigma_{\epsilon};\chi_{0},\pi_{0}\biggl\vert\frac{\delta S[\hat{\varphi},\hat{\varphi}^\dagger]}{\delta\hat{\varphi}^\dagger(g_I,\chi_{0})}\biggr\vert\sigma_{\epsilon};\chi_{0},\pi_{0}\right\rangle=0\,,
\end{align}
where $\ket{\sigma_{\epsilon};\chi_{0},\pi_{0}}$ is the isotropic CPS with wavefunction \eqref{eqn:wavefunctioncps} and with peaking function \eqref{eqn:peakingfunction}. The quantity $S$ is the GFT action. As we have mentioned at the beginning of Section \ref{subsec:reviewgft}, its form is chosen so that the GFT partition function expanded around the Fock vacuum matches the spin-foam model one wants to reproduce. Following \cite{Oriti:2016qtz}, this would be an EPRL Lorentzian model with a minimally coupled massless scalar field, described in terms of the $\text{SU}(2)$ projection of the Lorentz structures entering in the original definition of the model. The action includes a quadratic kinetic term and a quintic (in powers of the field operator) interaction term, $S=K+U+\bar{U}$.

For cosmological applications, there are typically two assumptions that are done on the GFT action. The first is that the classical field symmetries of the action of a minimally coupled massless scalar field (invariance under shift and reflection) are respected by the GFT action as well. This greatly simplifies the form of the interaction and kinetic terms, which read \cite{Oriti:2016qtz,Marchetti:2020umh}
\begin{align*}
    K&=\int\diff g_I\diff h_I\int \diff\chi\diff\chi'\,\bar{\varphi}(g_I,\chi)\\
    &\quad \times \mathcal{K}(g_I,h_I;(\chi-\chi')^2)\varphi(h_I,\chi')\,,\\
    U&=\int\diff\chi\int\left(\prod_{a=1}^5\diff g_I^a\right)\mathcal{U}(g_I^1,\dots,g_I^5)\prod_{a=1}^5\varphi(g_I^a,\chi)\,.
\end{align*}
The details about the EPRL model are encoded in the specific form of the kinetic and interaction kernels $\mathcal{K}$ and $\mathcal{U}$ above. In particular, it is $\mathcal{U}$ that carries information about the specific Lorentzian embedding of the theory. 

The second assumption usually made in cosmological applications, however, is that one is interested in a \virgolette{mescocopic regime} where these interactions are assumed to be negligible (though see \cite{Pithis:2016wzf,Pithis:2016cxg}, for some phenomenological studies including interactions). Under these two assumptions and imposing isotropy on the condensate wavefunction (see Subsection \ref{subsub:gftcondensatecosmology}), the above quantum equations of motion reduce to two equations for the modulus $\rho_j$ and the phase $\theta_j$ of the reduced wavefunction $\tilde{\sigma}_j\equiv \rho_j\exp[i\theta_j]$ of the CPS for each spin $j$ \cite{Marchetti:2020umh}, 
\begin{subequations}
\begin{align}
\label{eqn:fundamentalequation}
0&=\rho''_{j}(\chi_{0})-\frac{Q_{j}^2}{\rho_{j}^3(\chi_{0})}-\mu_{j}^2\rho_{j}(\chi_{0})\,,\\ \theta_{j}'(\chi_{0})&=\tilde{\pi}_{0}+\frac{Q_{j}}{\rho_{j}^2(\chi_{0})}\,,\label{eqn:phaseequation}
\end{align}
\end{subequations}
where 
\begin{subequations}
\begin{align}
\mu_{j}^2&=\frac{\pi_{0}^2}{\epsilon\pi_{0}^2-1}\left(\frac{2}{\epsilon\pi_{0}^2}-\frac{1}{\epsilon\pi_{0}^2-1}\right)+\frac{B_{j}}{A_{j}}\,,\\ \tilde{\pi}_{0}&=\frac{\pi_{0}}{\epsilon\pi_{0}^2-1}\,,
\end{align}
\end{subequations}
$Q_j$ are integration constants and \cite{Oriti:2016qtz}
\begin{align*}
    A_j&=\sum_{\vec{n},\iota}\left[\mathcal{K}^{(2)}\right]^{jjjj,\iota}_{n_1n_2n_3n_4}\bar{\mathcal{I}}^{jjjj\iota_+}_{n_1n_2n_3n_4}\mathcal{I}^{jjjj\iota_+}_{n_1n_2n_3n_4}\bar{\alpha}_j^{\iota}\alpha_j^{\iota},\\
    B_j&=-\sum_{\vec{n},\iota}\left[\mathcal{K}^{(0)}\right]^{jjjj,\iota}_{n_1n_2n_3n_4}\bar{\mathcal{I}}^{jjjj\iota_+}_{n_1n_2n_3n_4}\mathcal{I}^{jjjj\iota_+}_{n_1n_2n_3n_4}\bar{\alpha}_j^{\iota}\alpha_j^{\iota}\,.
\end{align*}
Here, $\mathcal{K}^{(2n)}$ denotes the $2n$-th derivative of the kineric kernel with respect to its scalar field argument evaluated at $0$, while $\alpha_j^\iota$ are determined by
\begin{equation*}
    \mathcal{I}^{jjjj,\iota_+}_{n_1n_2n_3n_4}=\sum_\iota \alpha^\iota_j\mathcal{I}^{jjjj,\iota}_{n_1n_2n_3n_4}\,.
\end{equation*}

Equation \eqref{eqn:fundamentalequation} can be immediately integrated once to obtain 
\begin{equation}\label{eqn:mathcalej}
\mathcal{E}_{j}=(\rho_{j}')^2+\frac{Q_{j}^2}{\rho_{j}^2}-\mu_{j}^2\rho_{j}^2\,,
\end{equation}
where the constants $\mathcal{E}_{j}$ are integration constants\footnote{It is interesting to notice that the above equation is equivalent to the equation of motion of a conformal particle \cite{deAlfaro1976} with a harmonic potential, which is a system characterized by a conformal symmetry. Since there are some interesting examples of systems whose dynamics can be can be mapped into a Friedmann one exactly in virtue of their conformal symmetry, (see e.g. \cite{BenAchour:2019ufa,Lidsey:2018byv}), the above equation alone would already suggest a connection between the effective mean field dynamics discussed here and a cosmological one.}. This equation can be then combined with equation \eqref{eqn:fundamentalequation} in order to obtain a linear equation in terms of $\rho_j^2(\chi_0)$. In fact, since $(\rho^2_{j})''=2(\rho_{j}')^2+2\rho_{j}\rho_{j}''$, combining equation \eqref{eqn:fundamentalequation} (multiplied by $\rho_{j}$) and equation \eqref{eqn:mathcalej}, we obtain
\begin{equation}
(\rho_{j}^2)''=2\left(\mathcal{E}_{j}+2\mu_{j}^2\rho_{j}^2\right)\,.
\end{equation}
The most general solution can be written as
\begin{equation}
\rho_{j}^2=-\frac{\mathcal{E}_{j}}{2\mu_{j}^2}+A_{j}e^{2\mu_{j}\chi_{0}}+B_{j}e^{-2\mu_{j}\chi_{0}}\,.
\end{equation}
Of course we can find some relations between the constants $A_{j}$ and $B_{j}$ and the constants of integration $Q_j$ and $\mathcal{E}_j$, so that we can choose a different way to parametrize the solution. Indeed, we first notice that since for $\chi_{0}\to+\infty$ the term with $A_{j}$ dominates, while for $\chi_{0}\to-\infty$ the term with $B_{j}$ dominates, this means that both $A_{j}$ and $B_{j}$ are non-negative. Then, by defining $\chi_{0,j}$ as the point at which $(\rho_{j}^2)'(\chi_{0,j})=0$, we see that
\begin{equation*}
\sqrt{A_{j}B_{j}}=\pm\frac{\sqrt{\mathcal{E}_{j}^2+4\mu_{j}^2Q_{j}^2}}{4\mu_{j}^2}\,,\qquad \sqrt{A_{j}/B_{j}}=e^{-2\mu_{j}\chi_{0,j}}\,.
\end{equation*}
Thus our solution becomes
\begin{subequations}\label{eqn:generalsolutiondifferentforms}
\begin{equation}
\rho_{j}^2=-\frac{\mathcal{E}_{j}}{2\mu_{j}^2}+\frac{\sqrt{\mathcal{E}_{j}^2+4\mu_{j}^2Q_{j}^2}}{2\mu_{j}^2}\cosh\left(2\mu_{j}(\chi_{0}-\chi_{0,j})\right)\,,
\end{equation}
where we have chosen only the positive solution because $\rho_{j}^2\ge 0$. Equivalently, we can write
\begin{equation}\label{eqn:generalsolution}
\rho_{j}^2=-\frac{\alpha_{j}}{2}+\frac{\sqrt{\alpha_{j}^2+4\beta_{j}^2}}{2}\cosh\left(2\mu_{j}(\chi_{0}-\chi_{0,j})\right),
\end{equation}
\end{subequations}
where we have defined
\begin{subequations}
\begin{equation}
\alpha_{j}\equiv \mathcal{E}_{j}/\mu_{j}^2\,,\qquad \beta_{j}^2\equiv Q_{j}^2/\mu_{j}^2\,.
\end{equation}
The solution is now only parametrized by $\mu_{j}$, $\alpha_{j}$, $\beta_{j}$, and $\chi_{0,j}$. This is our fundamental equation, representing a general solution of \eqref{eqn:fundamentalequation}. 

\

For the following discussion, it will be useful to derive some bounds on the modulus of the derivatives of $\rho_j^2$ divided by $\rho_j^2$ itself.  In order to study these bounds explicitly, it is helpful to define 
\begin{equation}
    x_j\equiv 2\mu_j(\chi_0-\chi_{0,j})\,,\qquad r_j\equiv \beta_j^2/\alpha_j^2\,.
\end{equation}
\end{subequations}
Then, denoting $[\rho_j^2]^{(n)}$ the $n$-th derivative of $\rho_j^2$ with respect to $\chi_0$, we have
\begin{align}\label{eqn:generalratio}
    \frac{\vert[\rho_j^2]^{(n)}\vert}{\rho_j^2}&=(2\mu_j)^n\frac{\sqrt{1+4r_j}     }{-\text{sgn}(\alpha_j)+\sqrt{1+4r_j}\cosh x_j}\nonumber\\
    &\quad\times\begin{cases}
    \sinh x_j\,,\qquad &n \text{ odd}\,.\\
    \cosh x_j\,,\qquad &n\text{ even}\,.
    \end{cases}
\end{align}
In the following sections we will discuss in more detail under which conditions the above states indeed implement a notion of relational dynamics, as defined in Subsection \ref{subsub:generalerd}.
\section{Averages and fluctuations: generalities}
In this section we compute expectation values of relevant operators in the effective relational GFT cosmology framework (i.e., the number operator $\hat{N}$, the volume operator $\hat{V}$, the momentum operator $\hat{\Pi}$, the Hamiltonian operator $\hat{H}$ and the massless scalar field operator $\hat{X}$), and their relative variances on CPS states, in order to characterize the behavior of the first moments of the relevant operators and with the ultimate purpose of trying to deduce some information about the impact of quantum fluctuations on the effective relational framework discussed so far (Section \ref{sec:quantumflucerd}).


We express these expectation values and relative variances in terms of the modulus of the reduced wavefunction only, possibly using equation \eqref{eqn:phaseequation} in order to trade any dependency on the phase of the reduced wavefunction for its modulus. In Section \ref{sec:explicitcomputations}, instead, we use the explicit solution \eqref{eqn:generalsolutiondifferentforms} to considerably simplify the equations obtained in the following two subsections. 
\subsection{Expectation value of relevant operators}\label{subsec:expvaluegeneral}
First, let us compute the expectation value of the relevant operators, whose definitions are reviewed below.
\paragraph*{Number and volume operators.}
The simple case of the number operator allows us to discuss the prototypical computation that we are going to perform frequently in the following. Its definition is \cite{Oriti:2016acw, Marchetti:2020umh}
\begin{equation}\label{eqn:numberoperator}
\hat{N}\equiv\int\diff\chi \int\diff g_{I}\,\hat{\varphi}^\dagger(g_{I},\chi)\varphi(g_{I},\chi)\,.
\end{equation}
Its expectation value on a isotropic CPS is therefore given by
\begin{equation*}
    N(\chi_0)\equiv \braket{\hat{N}}_{\sigma_\epsilon;\chi_0,\pi_0}=\sum_j\int\diff\chi \rho_j^2(\chi)\vert\eta_\epsilon(\chi-\chi_0;\pi_0)\vert^2\,.
\end{equation*}
In order to evaluate this quantity, one can expand the function $\rho_j^2$ around $\chi=\chi_0$, given the fact that the function $\eta_\epsilon$ is strongly peaked around that point. As a result, the relevant integral to be computed is
\begin{align*}
    \rho_j^2(\chi_0)\int\diff\chi&\vert\eta_\epsilon(\chi-\chi_0;\pi_0)\vert^2\\
    &\times\left[1+\sum_{n=1}^{\infty}\frac{(\chi-\chi_0)^n}{n!}\frac{[\rho_j^2]^{(n)}(\chi_0))}{\rho_j^2(\chi_0)}\right]
\end{align*}
By normalizing $\eta_\epsilon$ so that the integral of its modulus squared is unitary, we see that 
\begin{equation}
    \int\diff\chi\,(\chi-\chi_0)^{2m}\vert\eta_\epsilon(\chi-\chi_0;\pi_0)\vert^2=\epsilon^m\frac{(2m)!}{4^m m!}\,,
\end{equation}
giving zero, instead, for odd powers. In conclusion, one finds
\begin{equation}\label{eqn:completenumber}
N(\chi_0)=\sum_j    \rho_j^2(\chi_0)\left[1+\sum_{n=1}^{\infty}\frac{[\rho_j^2]^{(2n)}(\chi_0))}{\rho_j^2(\chi_0)}\frac{\epsilon^{n}}{4^{n} n!}\right]
\end{equation}
Similar computations hold for the volume operator, counting the volume contributions of all tetrahedra in a given GFT state and defined as \cite{Oriti:2016qtz,Marchetti:2020umh} 
\begin{equation}\label{eqn:volumeoperator}
\hat{V}=\int\diff\chi\int \diff g_I\diff g'_I\hat{\varphi}^\dagger(g_I,\chi)V(g_I,g'_I)\hat{\varphi}(g'_I,\chi)\,,
\end{equation}
in terms of matrix elements $V(g_I,g_I')$ of the first quantized volume operator in the group
representation. Indeed, one has \cite{Marchetti:2020umh}
\begin{align*}
    V(\chi_0)&\equiv \braket{\hat{V}}_{\sigma_\epsilon;\chi_0,\pi_0}\\&=\sum_jV_j\int\diff\chi \rho_j^2(\chi)\vert\eta_\epsilon(\chi-\chi_0;\pi_0)\vert^2\,,
\end{align*}
where $V_j$ represents the volume contribution of each equilateral tetrahedron whose faces have area determined by the quantum number $j$. The situation is the same as before, and one therefore concludes that
\begin{equation}\label{eqn:completevolume}
V(\chi_0)=\sum_jV_j    \rho_j^2(\chi_0)\left[1+\sum_{n=1}^{\infty}\frac{[\rho_j^2]^{(2n)}(\chi_0))}{\rho_j^2(\chi_0)}\frac{\epsilon^{n}}{4^{n} n!}\right].
\end{equation}
In particular, when higher order derivatives can be neglected, we notice that we can write 
\begin{equation}
    N(\chi_0)\simeq \sum_j\rho^2_j(\chi_0)\,,\qquad  V(\chi_0)\simeq \sum_jV_j\rho^2_j(\chi_0)\,,
\end{equation}
for the expectation value of the number and of the volume operator. We will see in Subsection \ref{subsec:explicitexpect} that this will be indeed the case when these quantities are evaluated on solutions of the dynamical equations and under some fairly general conditions on the parameters characterizing the dynamics. 
\paragraph*{Momentum and Hamiltonian operator.}
Similar results hold for the scalar field momentum and the hamiltonian operators. The effective\footnote{We remark that this is an \virgolette{effective} operator since its construction is always subject to a prior choice of appropriate states; in this case, CPSs (see \cite{Marchetti:2020umh} for a more detailed discussion).} Hamiltonian operator $\hat{H}$  can be defined as a Hermitean operator whose action on a CPS is given by \cite{Marchetti:2020umh}
\begin{widetext}
\begin{equation}\label{eqn:defh}
\hat{H}\ket{\sigma_{\epsilon};\chi_{0},\pi_{0}}\equiv -i\left(\frac{N'(\chi_{0})}{2}+\int\diff g_I\int\diff\chi\,\hat{\varphi}^\dagger(g_I,\chi)\partial_{\chi}\eta_{\epsilon}(\chi-\chi_{0},\pi_{0})\tilde{\sigma}(g_I,\chi)\right)\ket{\sigma_{\epsilon};\chi_{0},\pi_{0}}\,.
\end{equation}
\end{widetext}
Such an operator generates by construction translations with respect to $\chi_0$, and thus, in the regime in which the relational picture is well-defined, it is the operator generating relational evolution of expectation values of geometric operators.

Defining $\hat{\bar{H}}$ the operator whose action on the CPSs is given by the second term in the round brackets in equation \eqref{eqn:defh}, we see that its expectation value on an isotropic CPS is
\begin{align*}
\braket{\hat{\bar{H}}}_{\sigma_{\epsilon};\chi_{0},\pi_{0}}&=\pi_{0}\int\diff g_I\int\diff\chi\vert\eta_{\epsilon}(\chi-\chi_{0},\pi_{0})\vert^2\rho^2(g_I,\chi)\\&\quad+\frac{i}{2}N'(\chi_{0})
\end{align*}
By definition of $N(\chi_0)$ we can write
\begin{equation*}
\braket{\hat{\bar{H}}}_{\sigma_{\epsilon};\chi_{0},\pi_{0}}= \pi_{0}N(\chi_{0})+\frac{i}{2}N'(\chi_{0})\,,
\end{equation*}
so that, in conclusion we obtain, for $\hat{H}$,
\begin{equation}\label{eqn:generalexpvalueh}
\braket{\hat{H}}_{\sigma_{\epsilon};\chi_{0},\pi_{0}}= \braket{\hat{\bar{H}}}_{\sigma_{\epsilon};\chi_{0},\pi_{0}}-i\frac{N'(\chi_{0})}{2}= \pi_{0}N(\chi_{0})\,.
\end{equation}
The situation for the momentum operator is similar. By definition
\begin{equation}
    \label{eqn:momentumoperator}
\hat{\Pi}=\frac{1}{i}\int\diff g_I\int\diff\chi \left[\hat{\varphi}^\dagger(g_I,\chi)\left(\frac{\partial}{\partial\chi}\hat{\varphi}(g_I,\chi)\right)\right],
\end{equation}
and one has
\begin{align}\label{eqn:averagepi}
\braket{\hat{\Pi}}_{\sigma;\chi_{0},\pi_{0}}&=\frac{1}{i}\int\diff\chi\sum_{j}\overline{\sigma}_{\epsilon,j}(\chi;\chi_{0},\pi_{0})\partial_{\chi}\sigma_{\epsilon,j}(\chi;\chi_{0},\pi_{0})\nonumber\\&=\sum_{j}\int\diff\chi \rho_{j}^2(\chi)(\theta'_j(\chi)+\pi_{0})\vert\eta_{\epsilon}(\chi-\chi_{0};\pi_{0})\vert^2\nonumber\\&=\pi_{0}\!\!\left[\frac{1}{\epsilon\pi_{0}^2-1}+1\right]\!\sum_{j}\int\!\diff\chi \rho_{j}^2(\chi)\vert\eta_{\epsilon}(\chi-\chi_{0};\pi_{0})\vert^2\nonumber\\&\quad+\sum_{j}Q_{j}\nonumber\\&=  \pi_{0}\left(\frac{1}{\epsilon\pi_{0}^2-1}+1\right)N(\chi_{0})+\sum_{j}Q_{j}\,.
\end{align}
From this explicit form, we notice that the evaluation of both the expectation value of $\hat{H}$ and $\hat{\Pi}$ essentially reduces to an evaluation of the averaged number operator.

Now, in the approximation $\epsilon\pi_0^2\gg 1$, the two expectation values coincide, as required by the effective relational dynamics framework, for any values of $\chi_0$, as long as $\sum_jQ_j=0$. It is however interesting to notice that, as the number of GFT quanta increases, the impact of the second term above becomes decreasingly important. As a consequence, in the asymptotic regime $N\to\infty$, the equality between the expectation values of the Hamiltonian and the momentum operator is satisfied to any degree of accuracy required, regardless of the strict imposition of $\sum_j Q_j=0$. So, if one was interested only to the implementation of an effective relational framework in the thermodynamics regime $N\to\infty$, or at large condensate densities (which become large universe volumes), one might be formally dispensed from imposing the condition $\sum_j Q_j=0$. On the other hand, if one wants to describe mesoscopic intermediate regimes through the same formalism, then such a requirement needs to be imposed. In order to take into account these different possibilities, from now on we retain any $\sum_j Q_j$ term, setting it to zero only when necessary.
\paragraph*{Massless scalar field operator.}
The massless scalar field operator is defined as \cite{Oriti:2016qtz,Marchetti:2020umh}
\begin{equation}
    \label{eqn:scalarfieldoperator}
\hat{X}\equiv \int\diff g_I\int\diff \chi\,\chi\hat{\varphi}^\dagger(g_I,\chi)\hat{\varphi}(g_I,\chi)\,,
\end{equation}
so its expectation value on an isotropic CPS is given by
\begin{equation}
    \braket{\hat{X}}_{\sigma_\epsilon;\chi_0,\pi_0}=\sum_j\int\diff\chi\,\chi\rho_j^2(\chi)\vert\eta_\epsilon(\chi-\chi_0;\pi_0)\vert^2\,.
\end{equation}
Notice, however, that this operator is extensive (with respect to the GFT number of quanta, thus indirectly with respect to the universe volume), so it can not be directly related (not even in expectation value) to an intensive quantity such as the massless scalar field. This identification, however, can be meaningful for the rescaled operator $\hat{\chi}\equiv \hat{X}/\braket{\hat{N}}_{\sigma_\epsilon;\chi_0,\pi_0}$, at least when the average on a CPS $\ket{\sigma_\epsilon;\chi_0,\pi_0}$ is considered. 

The evaluation of $\braket{\hat{X}}_{\sigma_\epsilon;\chi_0,\pi_0}$ follows the lines described above: one has to first expand the integrand around $\chi_0$, and then integrates the expansion. As before, only terms with even number of derivatives survive the integration, so we can write the above quantity as 
\begin{align*}
 &\braket{\hat{X}}_{\sigma_\epsilon;\chi_0,\pi_0}=\chi_0\sum_j\rho_j^2(\chi_0)\int\diff\chi\,\vert\eta_\epsilon(\chi-\chi_0;\pi_0)\vert^2\Biggl\{1\\
 &\quad+\sum_{n=1}^\infty\left[\frac{[\rho_j^2]^{(2n)}(\chi_0)}{\rho_j^2(\chi_0)}+2n\frac{[\rho_j^2]^{(2n-1)}(\chi_0)}{\rho_j^2(\chi_0)\chi_0}\right]\frac{(\chi-\chi_0)^{2n}}{(2n)!}\Biggr\}\,. 
\end{align*}
As a result of the integration one obtains
\begin{align}\label{eqn:completemassless}
     &\braket{\hat{X}}_{\sigma_\epsilon;\chi_0,\pi_0}=\chi_0\sum_j\rho_j^2(\chi_0)\\&\quad\times\left\{1+\sum_{n=1}^\infty\left[\frac{[\rho_j^2]^{(2n)}(\chi_0)}{\rho_j^2(\chi_0)}+2n\frac{[\rho_j^2]^{(2n-1)}(\chi_0)}{\rho_j^2(\chi_0)\chi_0}\right]\frac{\epsilon^n}{4^n n!}\right\}\,.\nonumber
\end{align}
The first terms in squared brackets are the same that appear in the expectation value of the number operator. The second terms in square brackets are new. And these terms are in fact crucial: when they are not negligible, the expectation value $\braket{\hat{X}}_{\sigma_\epsilon;\chi_0,\pi_0}$ can not be written anymore as $\chi_0N(\chi_0)$, which means that the expectation value of the intrinsic massles scalar field operator $\hat{\chi}$ is not $\chi_0$ anymore. 

More precisely, in the most general case, by defining
\begin{equation}\label{eqn:deltax}
    \Delta X(\chi_0)\equiv \sum_j\sum_{n=1}^\infty 2n\frac{[\rho_j^2]^{(2n-1)}(\chi_0)}{\chi_0}\frac{\epsilon^n}{4^n n!}\,,
\end{equation}
we see that this leads to an expectation value of the \virgolette{intrinsic} massless scalar field operator of the form
\begin{align}
    \braket{\hat{\chi}}_{\sigma_\epsilon;\chi_0,\pi_0}&\equiv\frac{\braket{\hat{X}}_{\sigma_\epsilon;\chi_0,\pi_0}}{N(\chi_0)}\nonumber\\
    &=\chi_0\left(1+\Delta X(\chi_0)/N(\chi_0)\right)\,,
\end{align}
and when the second term satisfies $\vert\Delta X(\chi_0)\vert/N(\chi_0)\gtrsim 1$, the CPS parameter $\chi_0$ is not anymore the expectation value of the intrinsic massless scalar field operator and thus the $\chi_0$-evolution of averaged geometric operators cannot be interpreted as a relational dynamics. How larger is the second term with respect to $1$ depends clearly on two features of the state: (i) the impossibility of peaking precisely the clock value, i.e. sending ($\epsilon\to 0$), and (ii) the possibility for $N^{-1}$ to be large in the regime of small number of particles. 

Given that the reason why we can not take the limit $\epsilon\to 0$ is related to quantum fluctuations, and that, generally speaking, these are expected to become important when $N\ll 1$, the term $\Delta X/N$ encodes a first interplay between relationality and quantum properties of the clock. 

\subsection{Relative variances}
According to the requirements specified in Subsection \ref{subsub:generalerd}, it is fundamental to check the behavior of clock quantum fluctuations in order to understand whether the relational framework is truly realized at an effective level. 

Having done that, this analysis should be extended to all the relevant geometric operators in terms of which we write the emergent cosmological dynamics; this is true in particular for the volume operator, whose averaged dynamics was shown in \cite{Marchetti:2020umh} to reproduce, \virgolette{at late times} and under some additional assumptions, a Friedmann dynamics. In order for this \virgolette{late time regime} to be truly interpreted as a classical one, quantum fluctuations of the volume operator (and possibly also of the other physically interesting operators) should be negligible. 

We now proceed to study the behavior of these fluctuations, limiting ourselves here only to relative variances. The explicit computations of these quantum fluctuations can be found in Appendix \ref{app:explicitfluctuations}.
\paragraph*{Number operator.}
As before, we start from the number operator. Its relative variance can be easily found to be
\begin{subequations}
\begin{equation}
\delta^2_{N}= N^{-1}(\chi_{0})\,,
\end{equation}
thus decreasing as the number of atoms of space increases, as expected. When the lowest order saddle point approximation is justified, one can write the above expression as
\begin{equation}
\delta^2_{N}\simeq \biggl[\sum_j\rho_j^2(\chi_0)\biggr]^{-1}\,.
\end{equation}
\end{subequations}
\paragraph*{Volume operator.}
For the volume operator the computations are similar. One finds
\begin{subequations}
\begin{equation}
    \delta^2_V=\frac{\sum_jV_j^2\rho_j^2\left[1+\sum_{n=1}^{\infty}\frac{[\rho_j^2]^{(2n)}(\chi_0))}{\rho_j^2(\chi_0)}\frac{\epsilon^{n}}{4^{n} (n)!}\right]}{\left\{\sum_jV_j\rho_j^2\left[1+\sum_{n=1}^{\infty}\frac{[\rho_j^2]^{(2n)}(\chi_0))}{\rho_j^2(\chi_0)}\frac{\epsilon^n}{4^{n} (n)!}\right]\right\}^2}\,.
\end{equation}
If one can neglect higher order derivatives, then
\begin{equation}
    \delta^2_{V}\simeq \frac{\sum_{j}V_{j}^2\rho_{j}^2}{(\sum_{j}V_{j}\rho_{j}^2)^2}\,.
\end{equation}
\end{subequations}
\paragraph*{Hamiltonian operator.}
The relative variance of the Hamiltonian operator, instead, is given by
\begin{equation}
\delta^2_{H}\simeq N^{-1}(\chi_{0})\left[1+\left(2\epsilon\pi_{0}^2\right)^{-1}\right]\simeq N^{-1}(\chi_0)\,,
\end{equation}
which is under control in the regime $\epsilon\pi_{0}^2\gg 1$ with a large number of GFT quanta and, in this limit, behaves as the relative variance of the number operator.
\paragraph*{Momentum operator.}
Next, we discuss the variance of the momentum operator. The computations are a little more involved, but under the assumption that $\epsilon\pi_0^2\gg 1$ one finds
\begin{align*}
    &\delta^2_\Pi=\frac{1}{(\pi_0N+\sum_jQ_j)^2}\Biggl\{\sum_j\left(Q_j+\mathcal{E}_j\right)\\
    &\quad+\sum_j\left[\frac{1}{\epsilon}+\mu_j^2+\pi_0^2\right]\rho_j^2(\chi_0)\left[1+\sum_{n=1}^\infty\frac{[\rho_j^2]^{(2n)}(\chi_0)}{\rho_j^2(\chi_0)}\frac{\epsilon^n}{n!4^n}\right] \\
    &\quad-\sum_j\frac{\rho_j^2(\chi_0)}{2\epsilon}\left(1+\sum_{n=1}^\infty\frac{[\rho_j^2]^{(2n)}(\chi_0)}{\rho_j^2(\chi_0)}\frac{\epsilon^n}{n!4^n}(2n+1)\right)\Biggr\} \, .
\end{align*}
From the explicit form of $\delta_\Pi^2$ above we deduce that: 
\begin{itemize}
    \item in the formal limit $\epsilon\to 0$, $\delta_\Pi^2\to\infty$, i.e. that the system is subject to arbitrarily large quantum fluctuations. This is of course a consequence of the Heisenberg uncertainty principle when \virgolette{clock time} localization of the condensate wavefunction is enhanced;
    \item by taking the limit\footnote{As we have mentioned above, the result above was obtained using the condition $\epsilon\pi_0^2\gg 1$ (see equation \eqref{eqn:approxpi}), which is certainly not justified in this case. However, one can explicitly check, by using the full result in terms of $\tilde{\pi_0}$ and $\pi_0$, that the conclusions below are still valid.} $\pi_0\to 0$, the expectation value of $\hat{\Pi}$ on a CPS becomes $\sum_j Q_j$, while one can see that $\delta \Pi^2_{\sigma_\epsilon;\chi_0,\pi_0}$ has a contribution growing essentially as $N$ (see Appendix \ref{app:explicitfluctuations}). In this case, therefore, quantum fluctuations become extremely large in the $N\to\infty$ regime, which is certainly an undesired feature, since we expect that in this limit some kind of semi-classical spacetime structure is recovered. On the other hand, as we will see below, when condition \eqref{eqn:epi} is assumed, in the limit $N\to\infty$ quantum fluctuations are suppressed.
\end{itemize}
\paragraph*{Massless scalar field operator.}
Finally, we discuss the variance of the massless scalar field operator. Its quantum fluctuations are given by
\begin{align}\label{eqn:deltaxsquared}
   \delta X^2_{\sigma_\epsilon;\chi_0,\pi_0}&
   =\chi_0^2\sum_j\rho_j^2(\chi_0)\biggl\{1+\sum_{n=1}^\infty\biggl[\frac{[\rho_j^2]^{(2n)}(\chi_0)}{\rho_j^2(\chi_0)}\nonumber\\
   &\quad+4n \frac{[\rho_j^2]^{(2n-1)}(\chi_0)}{\chi_0\rho_j^2(\chi_0)}\nonumber\\
   &\quad+2n(2n-1)\frac{[\rho_j^2]^{(2n-2)}(\chi_0)}{\chi_0^2\rho_j^2(\chi_0)}\biggr]\frac{\epsilon^n}{4^n n!}\biggr\},
\end{align}
which, once divided by \eqref{eqn:completemassless} squared, gives the relative variance of $\hat{\chi}$. Notice, in the above equation, that even though the coefficients in the square brackets grow as $n$ for the second term and as $n^2$ for the third one, the behavior of the overall coefficients of these terms (i.e., by taking into account also the factor $(4^n n!)^{-1}$) is decreasing with $n$. As we will see in the next section, this implies that in the evaluation of this variance on the specific solutions \eqref{eqn:generalsolution} it is enough to consider the lowest non-trivial order. The only difference with respect to the expectation value of the massless scalar field, is that in this case odd and even derivatives of the $\rho_j^2$ function are at the same perturbative order. In particular, for $n=1$ the last term becomes dominant when $\epsilon/(2\chi_0)^2\gg 1$, i.e., when $\vert\chi_0\vert \ll \sqrt{\epsilon/2}$. Now, suppose that $ \vert\pi_0\vert^{-1}\ll\vert\chi_0\vert\ll\sqrt{\epsilon/2}$ (this region is allowed because $\epsilon\pi_0^2\gg 1$), so that the computations carried out for the expectation value are still valid, but this last term is indeed important in the evaluation of the fluctuations. We see that this $n=1$ term gives a contribution to the relative variance of the form
\begin{equation*}
    \frac{\epsilon}{2\chi_0^2}\frac{\sum_j\rho_j^2(\chi_0)}{\left[\sum_j\rho_j^2(\chi_0)^2\right]^2}\simeq  \frac{\epsilon}{2\chi_0^2}N^{-1}(\chi_0)\,.
\end{equation*}
The prefactor on the right-hand-side is by assumption large, but it can be suppressed by the factor $N^{-1}(\chi_0)$, assuming it is large enough. 

So, already from this example we can deduce that, in the limit of arbitrarily large $N$, the only point which has to formally be excluded from the analysis because clock fluctuations become too large is $\chi_0=0$. In this point the prefactor is formally divergent, regardless of any large value of $N$ we are considering. This is of course a consequence of using relative variances: if we are interested in the physics at $\chi_0=0$, as already argued, e.g., in \cite{Ashtekar:2005dm}, one should set a precise threshold on $\delta X^2$, rather than using relative variances.
\section{Averages and fluctuations: explicit evaluation}\label{sec:explicitcomputations}
Having obtained the expectation values and the relative variances of relevant operators in the effective relation GFT cosmology frameowork in terms of the modulus of the reduced wavefunction (and possibily of its derivatives), we can now further simplify the obtained expressions by means of equation \eqref{eqn:generalsolution}.
\subsection{Expectation values}\label{subsec:explicitexpect}
The explicit evaluation of the expectation values of operators, as shown above, involves an infinite number of derivatives of the modulus of the reduced wavefunction. However, it is interesting to notice that
\begin{align}\label{eqn:evederivativesapprox}
    1+\frac{[\rho_j^2]''}{\rho_j^2}\frac{\epsilon}{4}&=\frac{-\text{sgn}(\alpha_j)+\sqrt{1+4r_j}\cosh x_j(1+\epsilon\mu_j^2)}{-\text{sgn}(\alpha_j)+\sqrt{1+4r_j}\cosh x_j}\nonumber\\&\simeq 1\,,
\end{align}
since
\begin{equation}\label{eqn:smallepsilonmu}
\mu_{j}^2\epsilon=\frac{\epsilon\pi_{0}^2}{\epsilon\pi_{0}^2-1}\left(\frac{2}{\epsilon\pi_{0}^2}-\frac{1}{\epsilon \pi_{0}^2-1}\right)+\epsilon\frac{B_{j}}{A_{j}}\ll 1 \, ,
\end{equation}
under our working assumption $\epsilon\pi_{0}^2\gg 1$ and by further assuming\footnote{\label{footnote:detailsgft}Notice that under this assumption, which seems natural given the smallness of $\epsilon$ required by the CPS construction, the details of the underlying GFT model become effectively unimportant for the derivation of the results discussed below.} $\vert B_{j}/A_{j}\vert\ll \epsilon^{-1}$. Moreover, since, in general, one has 
\begin{equation}\label{eqn:derivativehierarchy}
    \frac{[\rho^2]^{(n+2)}}{[\rho^2]^{(n)}}=4\mu_j^2\,,\qquad n\ge 1\,,
\end{equation}
we see that terms with $n>1$ are negligible with respect to the $n=1$ term, thus implying that \emph{all} even derivatives in the sums \eqref{eqn:completenumber} and \eqref{eqn:completevolume} can be neglected. As a result, we can write
\begin{equation}\label{eqn:approximatenv}
    N(\chi_0)\simeq \sum_j\rho^2_j(\chi_0)\,,\qquad V(\chi_0)\simeq \sum_jV_j\rho^2_j(\chi_0)\,,
\end{equation}
the first equation above also determining the expectation values of $\hat{\Pi}$ and $\hat{H}$, according to equations \eqref{eqn:generalexpvalueh} and 
 \eqref{eqn:averagepi}.
 
Similarly, the natural hierarchy of derivatives obtained from equations \eqref{eqn:derivativehierarchy} and \eqref{eqn:smallepsilonmu} is present also in in the sum over $n$ in equation \eqref{eqn:deltax}. For the same reasons as above, therefore, we can write
\begin{equation*}
    \Delta X(\chi_0)\simeq \sum_j\frac{[\rho_j^2]'(\chi_0)}{\chi_0}\frac{\epsilon}{2}\,,
\end{equation*}
so that
\begin{equation}\label{eqn:deltaxnsimp}
    \frac{\vert \Delta X\vert}{N}\simeq \biggl\vert\sum_j\frac{[\rho_j^2]'(\chi_0)}{\chi_0}\frac{\epsilon}{2}\biggr\vert\biggl[ \sum_j\rho^2_j(\chi_0)\biggr]^{-1}.
\end{equation}
However, determining whether these higher derivative terms become important in the expectation values of operators like $\hat{N}$, $\hat{V}$, $\hat{\Pi}$ and $\hat{H}$ is quite different from determining whether $\Delta X/N$ is important in the expectation value of $\hat{\chi}$, basically because, as we can clearly see from the above expression, $\Delta X$ involves odd derivatives and it explicitly depends on $\chi_0$. 

As we have already mentioned, the smallness of the factor $\Delta X/N$ is crucial for a consistent interpretation of $\chi_0$ as the expectation value of the massless scalar field $\hat{\chi}$, to be used in a relational picture. In general, whenever $\vert\Delta X\vert/N\ll 1$, such an interpretation is allowed. 

Whether this condition is actually satisfied, though, drastically depends on the properties of the solution $\rho_j^2$ for each $j$ and hence on the precise set of free parameters $\{\alpha_j,\beta_j,\chi_{0,j}\}$. It is obvious from equation \eqref{eqn:completemassless} together with the above expression, that as long as
\begin{equation}\label{eqn:smalldeltaxcondition}
    1+\frac{[\rho_j^2]'}{\rho_j^2\chi_0}\frac{\epsilon}{2}\simeq 1\,,
\end{equation}
this interpretation is allowed. Following the same steps of \eqref{eqn:evederivativesapprox}, we see that the above condition is satisfied as long as 
\begin{equation*}
    \epsilon\mu_j^2\frac{\vert \tanh x_j\vert}{\vert x_j+x_j^o\vert}\ll 1\,, \qquad \forall j
\end{equation*}
where $x_j^o\equiv 2\mu_j\chi_{0,j}$, and where we have neglected an unimportant factor $2$. This condition is certainly satisfied in two simple (though interesting) cases:
\begin{enumerate}
    \item First, since $\vert \tanh x_j\vert\le 1$, we see that when $\vert x_j+x_j^o\vert\equiv 2\mu_j\vert\chi_0\vert\gg (\epsilon\mu_j^2)^{-1}$, i.e., again, neglecting unimportant factors $2$, when
    \begin{equation}\label{eqn:simplestconditionchi0}
        \vert\chi_0\vert\gg \epsilon\mu_j\sim \pi_0^{-1}\,,
    \end{equation}
    condition \eqref{eqn:smalldeltaxcondition} is actually satisfied. Notice also that since $\pi_0^{-1}\ll\sqrt{\epsilon}$, by requiring $\vert\chi_0\vert\gg\sqrt{\epsilon}$ the above condition is also satisfied. It is interesting to notice that $\sqrt{\epsilon}$ actually quantifies the impossibility to perfectly localize the condensate wavefunction around $\chi_0$. If $\chi_0$ is of order or smaller than this quantity, it is clear that any desired localization property is lost in this irreducible uncertainty. 
    \item Second, notice that if all the $x_j^o\ge 0$ (resp. $x_j^o\le 0$) the above condition is always satisfied for all $\chi_0\ge 0$ (resp. for all $\chi_0\le 0$). In the case only a single spin is considered, say $j_o$, this means that the evolution of the modulus of the condensate wavefunction with respect to $\chi_0$ can be interpreted as an evolution with respect to the expectation value of $\hat{\chi}$ from the minimum of the former, at $x_{j_o}^o\ge 0$ (resp. $x_{j_o}^o\le 0$) to arbitrarily large positive (negative) values of $\chi_0$. We will discuss this single spin case in more detail in Subsection \ref{subsub:singlespin}.
\end{enumerate}
More generally, instead, the value of $h_j(x_j)\equiv \vert\tanh x_j\vert/\vert x_j+x_j^o\vert$ is determined by two scales: $ x_j+x_j^o\equiv \kappa^{(1)}_j\equiv 2\mu_j\chi_0$, and $x_j^o/\kappa^{(1)}_j\equiv \kappa^{(2)}_j$. These two quantities acquire a clear physical meaning in a single spin scenario with $j=j_o$ discussed in Subsection \ref{subsub:singlespin}. In that case, $\kappa_{j_o}^{(1)}$ basically measures the amount of evolution experienced by $\rho_{j_o}^2$ from $\chi_0=0$, while $\kappa_{j_o}^{(2)}$ measures how large is the amount of evolution elapsed since $\chi_0=0$ with respect to the moment at which $\rho_{j_o}^2$ has reached its minimum. Since only a single spin is excited, the expectation value of the volume operator and $\rho_{j_o}^2$ are in a one-to-one correspondence (see equation \eqref{eqn:approximatenv}), which gives to the above statements about $\kappa_k^{(1)}$ and $\kappa_j^{(2)}$ a straightforward physical meaning.   

Of course, the desired condition \eqref{eqn:smalldeltaxcondition} is satisfied for $\vert \kappa^{(1)}_j\vert\gtrsim 1$ (late evolution for $\rho_j^2$) or for $\vert\kappa^{(1)}_j\vert\ll 1$ and $\vert\kappa^{(2)}_j\vert\lesssim 1$ (early evolution for $\rho_j^2$, but still later than when the minimum of $\rho_j^2$ happened), as reviewed in Table \ref{tab:k1k2}. 
In the remaining cases, $\vert\kappa^{(1)}_j\vert\ll 1$ and $\vert\kappa^{(2)}_j\vert\gg 1$ (very early evolution for $\rho_j^2$), instead, we have 
\begin{equation*}
    h_j(x_i)\sim \begin{cases}
    \vert\kappa^{(1)}_j\vert^{-1}\,,\quad &\vert\kappa^{(1)}_j\vert\vert\kappa^{(2)}_j\vert\gtrsim 1\\
    \vert\kappa^{(2)}_j\vert\,,\quad &\vert\kappa^{(1)}_j\vert\vert\kappa^{(2)}_j\vert\ll 1
    \end{cases}\,,
\end{equation*}
so, while in the first case the condition $\epsilon\mu_j^2 h_j\ll 1$ becomes the condition already encountered in \eqref{eqn:simplestconditionchi0}, $\vert\chi_0\vert \gg\epsilon\mu_j$, in the second case the situation is different. We see that when
\begin{equation*}
    \epsilon\mu_j^2\times\begin{cases}
    \vert\kappa^{(1)}_j\vert^{-1}\ll1\,,\quad &\vert\kappa^{(1)}_j\vert\vert\kappa^{(2)}_j\vert\gtrsim 1\\
    \vert\kappa^{(2)}_j\vert\ll 1\,,\quad &\vert\kappa^{(1)}_j\vert\vert\kappa^{(2)}_j\vert\ll 1
    \end{cases}\,,
\end{equation*}
for all $j$s, then we can write $\braket{\hat{X}}_{\sigma_\epsilon;\chi_0,\pi_0}\simeq \chi_0\sum_j\rho_j^2(\chi_0)\simeq \chi_0N(\chi_0)$, and conclude that $\chi_0$ is indeed the expectation value of the intrinsic massless scalar field operator $\hat{\chi}$. See Tables \ref{tab:k1k2} and \ref{tab:k1k2details} for a summary of the results.
\begin{table}
\centering
\begin{tabular}{|c||*{3}{c|}}\hline
\diagbox{$\bigl\vert\kappa_j^{(2)}\bigr\vert$}{$\bigl\vert\kappa_j^{(1)}\bigr\vert$}
&\makebox[3em]{$\ll 1$}&\makebox[3em]{$\sim 1$}&\makebox[3em]{$\gg 1$}\\\hline\hline
$\ll 1$ & $\checkmark$ & $\checkmark$  & $\checkmark$ \\\hline
$\sim 1$ &$\checkmark$ & $\checkmark$  & $\checkmark$ \\\hline
$\gg 1$ & ? & $\checkmark$  & $\checkmark$ \\\hline
\end{tabular}
\caption{Validity of the condition \eqref{eqn:smalldeltaxcondition} depending on the scales $\kappa_j^{(1)}$ and $\kappa_j^{(2)}$. The case $\vert\kappa_j^{(1)}\vert\ll 1$ and  $\vert\kappa_j^{(2)}\vert\gg 1$ is studied in Table \ref{tab:k1k2details} below.}
\label{tab:k1k2}
\end{table}
\begin{table}
\centering
\begin{tabular}{|c||*{2}{c|}}\hline
\diagbox[width=5cm]{$(\epsilon \mu_j)^2/\bigl\vert\kappa_j^{(1)}\bigr\vert$}{$\bigl\vert\kappa_j^{(1)}\kappa_j^{(2)}\bigr\vert$}
&\makebox[3em]{$\ll 1$}&\makebox[3em]{$\gtrsim 1$}\\\hline\hline
$\ll 1$ & $\text{\sffamily X}$ &  $\checkmark$  \\\hline
$\gtrsim 1$ & $\text{\sffamily X}$ &  $\text{\sffamily X}$ \\\hline
\end{tabular}\\
\vspace{3mm}
\begin{tabular}{|c||*{2}{c|}}\hline
\diagbox[width=5cm]{$(\epsilon \mu_j)^2\bigl\vert\kappa_j^{(2)}\bigr\vert$}{$\bigl\vert\kappa_j^{(1)}\kappa_j^{(2)}\bigr\vert$}
&\makebox[3em]{$\ll 1$}&\makebox[3em]{$\gtrsim 1$}\\\hline\hline
$\ll 1$ &  $\checkmark$ & $\text{\sffamily X}$  \\\hline
$\gtrsim 1$ & $\text{\sffamily X}$ &  $\text{\sffamily X}$  \\\hline
\end{tabular}
\caption{Validity of the condition \eqref{eqn:smalldeltaxcondition} depending on the scales $\kappa_j^{(1)}$ and $\kappa_j^{(2)}$, assuming $\vert\kappa_j^{(1)}\vert\ll 1$ and  $\vert\kappa_j^{(2)}\vert\gg 1$.}
\label{tab:k1k2details}
\end{table}

\subsection{Fluctuations}
The arguments exposed above can be used straightforwardly to compute relative variances of operators. 
\paragraph*{Number, Hamiltonian and Volume.}
For the relative variances of the number, Hamiltonian and volume operators, we have
\begin{subequations}\label{eqn:variances}
\begin{align}\label{eqn:variancestrivial}
    \delta^2_H&\simeq \delta^2_N=N^{-1}\simeq \biggl[\sum_j\rho_j^2\biggr]^{-1}\\\label{eqn:variancevolumefinal}
    \delta^2_V&\simeq \sum_j V^2_j\rho_j^2/\biggl[\sum_j V_j\rho_j^2\biggr]^2\,.
\end{align}
\paragraph*{Momentum.}
For the momentum operator, given that we require $\epsilon\mu_j^2\ll 1$, we can safely only retain the first terms of the expansions appearing in \eqref{eqn:secondsimplifydeltapi}. Moreover, since $\mu_j^2\sim (\epsilon\pi_0)^{-2}$, $\mu_j^2\pi_0^2\sim (\epsilon\pi_0^2)^{-2}$, and since $\epsilon\pi_0^2\gg 1$, both the first two terms in squared brackets in the first line of \eqref{eqn:secondsimplifydeltapi}, as well as the whole first term in the second line of equation \eqref{eqn:secondsimplifydeltapi} are negligible with respect to the term
\begin{equation*}
    \pi_0^2\sum_{j}\mathcal{N}_{\epsilon}^2\int\diff\chi\rho_{j}^2e^{-\frac{(\chi-\chi_{0})^2}{\epsilon}}=\pi_0^2N(\chi_0)\,.
\end{equation*}
As a result, we finally have
\begin{subequations}
\begin{equation*}\label{eqn:deltapisquared}
    \delta \Pi^2_{\sigma_\epsilon;\chi_0,\pi_0}\simeq \pi_0^2N(\chi_0)+\sum_j\left(\mathcal{E}_j+Q_j\right)\,.
\end{equation*}
Now, we recall that in order to have an identification between the first moments of the Hamiltonian and the momentum operator one needs to have either $\sum_jQ_j=0$ or to be in the asymptotic limit in which $\sum_jQ_j$ is negligible with respect to $\pi_0 N$. Since $\vert\pi_0\vert>1$ this implies that when this identification is true, then we can also neglect the $\sum_jQ_j$ term in the above equation. As a consequence, we have a relative variance
\begin{align*}
    \delta^2_\Pi&\simeq N^{-1}(\chi_0)+N^{-2}(\chi_0)\frac{\sum_j\mathcal{E}_j}{\pi_0^2}\nonumber\\
    &\simeq N^{-1}+N^{-2}\sum_j\mu_j^2\alpha_j\,.
\end{align*}
\end{subequations}
So, we see that the first term of the relative variance behaves as $\sigma^2_N$, while the second is new, and because of its behavior $\sim N^{-2}$ it might become dominant in the regime in which $N\ll 1$.

Also, let us notice that $\delta \Pi^2_{\sigma_\epsilon;\chi_0,\pi_0}$ is indeed always positive under our assumptions. In fact, we see that we can write
\begin{align*}
    \delta \Pi^2_{\sigma_\epsilon;\chi_0,\pi_0}=\pi_0^2\sum_j\rho_j^2\Biggl[&1+\frac{\sum_{j\in P}\alpha_j(\mu_j^2/\pi_0^2)}{\sum_j\rho_j^2}\\&-\frac{\sum_{j\in N}\vert\alpha_j\vert(\mu_j^2/\pi_0^2)}{\sum_j\rho_j^2}\Biggr],
\end{align*}
where $P\equiv \{j\in J\mid \alpha_j\ge 0\}$, and $N\equiv J-P$, $J$ being the total set of spins over which the sum is performed\footnote{We will formally assume that the set $J$ is finite, either because there is an explicit cut-off $\Lambda $ on the allowed spins, or because, after a certain spin $\Lambda$ on, all the $\rho_j$s become dynamically subdominant.}. 

Let us estimate how large is the last term in square brackets. Since $\mu_j^2/\pi_0^2\sim (\epsilon\pi_0^2)^{-2}$, we can bring it out of the sum and just study
\begin{equation*}
    \frac{\sum_{j\in N}\vert\alpha_j\vert}{\sum_j\rho_j^2}\le\frac{\sum_{j\in N}\vert\alpha_j\vert}{\sum_{j\in N}\rho_j^2}\le 1\,,
\end{equation*}
since, for each $j \in N$,
\begin{equation*}
    \rho_j^2=\vert\alpha_j\vert(1+\sqrt{1+4r_j}\cosh x_j)/2\ge \vert\alpha_j\vert\,.
\end{equation*}
Thus, generally speaking, we have
\begin{equation*}
    \delta \Pi^2_{\sigma_\epsilon;\chi_0,\pi_0}\ge\pi_0^2\sum_j\rho_j^2\Biggl[1+\frac{\sum_{j\in P}\alpha_j(\mu_j^2/\pi_0^2)}{\sum_j\rho_j^2}-\frac{1}{(\epsilon\pi_0^2)^2}\Biggr],
\end{equation*}
and the right-hand-side is of course positive because $(\epsilon\pi_0^2)\gg 1$. Moreover, since
\begin{equation*}
    \delta \Pi^2_{\sigma_\epsilon;\chi_0,\pi_0}\le\pi_0^2\sum_j\rho_j^2\Biggl[1+\frac{\sum_{j\in P}\alpha_j(\mu_j^2/\pi_0^2)}{\sum_j\rho_j^2}\Biggr],
\end{equation*}
we see that in this limit we can approximately write
\begin{equation*}
    \delta \Pi^2_{\sigma_\epsilon;\chi_0,\pi_0}\simeq \pi_0^2N(\chi_0)\left[1+N^{-1}(\chi_0)\sum_{j\in P}\alpha_j(\mu_j^2/\pi_0^2)\right],
\end{equation*}
so that the relative variance becomes 
\begin{equation}\label{eqn:finalsigmapi}
    \delta^2_\Pi\simeq N^{-1}(\chi_0)+N^{-2}(\chi_0)\sum_{j\in P}\alpha_j(\mu_j^2/\pi_0^2) \, .
\end{equation}

\paragraph*{Massless scalar field.}
Instead, about the relative variance of the massless scalar field operator, we have
\begin{align}\label{eqn:variancechi}
    \delta^2_\chi&\simeq \frac{\epsilon}{2N\chi_0^2}\frac{1}{(1+\Delta X/N)^2}+\frac{N+2\Delta X}{(N+\Delta X)^2}\\
    &\lesssim\frac{1}{N}\left(1+ \frac{\epsilon}{2\chi_0^2}\frac{1}{(1+\Delta X/N)^2}\right)\nonumber
\end{align}
\end{subequations}
   \setcounter{equation}{48}
Let us make two remarks about this quantity:
\begin{enumerate}
    \item First, in order for \eqref{eqn:deltaxsquared} to be non-negative, we need to impose that
    \begin{equation*}
        \Delta X/N\ge -1/2-\epsilon/(4\chi_0^2)\,.
    \end{equation*}
    Contrarily to what happens for the momentum operator, this is actually a feature that we must impose \virgolette{by hand} on our solutions. We will assume it to be true from now on.
    \item Second, the divergence in the above variance at the point $\Delta X/N=1$ is again due to our choice of using relative variances, and, as already mentioned above, a more careful choice would be to define an appropriate threshold on the quantity $\delta X^2_{\sigma_\epsilon;\chi_0,\pi_0}$  \cite{Ashtekar:2004eh,Marchetti:2020umh}. The precise identification of this threshold is usually demanded to observational constraints, which are not available in our case. As a consequence, we just consider relative variances, and avoid the point in which they might diverge.  
\end{enumerate}
The scaling of all the relative variances \eqref{eqn:variances} is essentially determined by $N\simeq \sum_j\rho_j^2$, so it is interesting to study separately situations in which $N\gg 1$ and $N\lesssim 1$.
\subsubsection{Large number of GFT quanta}\label{subsub:largenumber}
We will first consider the case of large number of GFT quanta, $N\gg 1$. Generally speaking, this situation might be realized in two different ways:
\begin{enumerate}
    \item There exists at least one of the $\rho_j^2$s which is much larger than one.
    \item All the $\rho_j^2$s are $\lesssim 1$, but their sum is still much larger than $1$. 
\end{enumerate}
While for the number and the Hamiltonian operators variances are smaller than one by assumption, for the momentum, the volume and the massless scalar field operator, the situation is more complicated, so it is useful to discuss them by distinguishing between the two cases above. 
\paragraph*{First case.} When at least one of the $\rho_j^2$s is much larger than one, it is useful to distinguish between two sets, $L\equiv \{j\in J\mid \rho_j^2\gg 1\}$, $S\equiv J-L$. 
\begin{description}
\item[$\delta^2_\Pi$] While the first term of the variance of the momentum operator is certainly much smaller than one, in order to evaluate the second term one should know exactly the values of all the $\alpha_j$s for each $j\in P$. Nonetheless, since $\mu_j^2/\pi_0^2\sim (\epsilon\pi_0^2)^{-2}$, the second term is actually neglgible as long as 
\begin{equation*}
    \sum_{j\in P}\alpha_j\ll [N(\chi_0)/(\epsilon\pi_0^2)]^2\,,
\end{equation*}
which is certainly satisfied for a large class of initial conditions, given the large value of the right-hand-side. 

For instance, notice that for $r_j\gtrsim 1$ for each $j\in P$, we have that
\begin{equation*}
    \sum_{j\in P}\alpha_j\lesssim \sum_{j\in P}\rho_j^2\lesssim N\ll N^2(\chi_0)/(\epsilon\pi_0^2)^2\,.
\end{equation*}
Also, notice that when $P=\emptyset$, $\sigma^2_\Pi\sim N^{-1}\ll 1$ under our assumptions.
\item[$\delta^2_V$] About the volume operator, we have the following set of inequalities:
\begin{align*}
    \delta^2_V&= \frac{\sum_{j\in L}V_j^2\rho_j^2}{\left(\sum_jV_j\rho_j^2\right)^2}+\frac{\sum_{j\in S}V_j^2\rho_j^2}{\left(\sum_jV_j\rho_j^2\right)^2}\\
    &\ll\frac{\sum_{j\in L}V_j^2\rho_j^4}{\left(\sum_jV_j\rho_j^2\right)^2}+\frac{\sum_{j\in S}V_j^2\rho_j^2}{\left(\sum_jV_j\rho_j^2\right)^2}\\&\ll 1+\frac{\sum_{j\in S}V_j^2\rho_j^2}{\left(\sum_jV_j\rho_j^2\right)^2}\le 1+\frac{\sum_{j\in S}V_j^2\rho_j^2}{\left(\sum_{j\in L}V_j\rho_j^2\right)^2}\\
    &\ll 1+(V^2)_S/(V^2)_L,
\end{align*}
where $(V^2)_{S,L}=\sum_{j\in (S,L)}V_j^2$. For $(V^2)_S/(V^2)_L\sim 1$, the variance of the volume operator is always much smaller than a quantity of order $1$ and thus it is negligible. 

In particular, when $L=J$, it follows that $\delta^2_V\ll 1$. Then the volume behaves classically, since \emph{all} the moments of the volume operator are negligible, as one can easily see by following the same steps taken for the variance. 
\item[$\delta^2_\chi$] As for the massless scalar field operator, we see that, when $\vert \Delta X\vert/N\ll 1$, the relative variance is negligible as long as $\chi_0^2\gg \epsilon/N$ (neglecting unimportant factors $2$). When, on the other hand this quantity is of order $1$, fluctuations on the massless scalar field operator might become important. Again, by definition, the point $\chi_0=0$ is a point where relative quantum fluctuations become uncontrollable. 

On the other hand, let us consider the situation in which $\vert \Delta X\vert/N\gtrsim 1$ (though not very close to $ \Delta X/N=-1$ leading to the unphysical singularity on the relative variance discussed above). In such a situation, we can consider the factor $1/\vert 1+\Delta X/N\vert\equiv1/ \lambda\lesssim 1$. The condition for having small fluctuations in this case becomes
\begin{equation}\label{eqn:largensmallfluctx}
    \vert\chi_0\vert \gg \sqrt{\epsilon/(\lambda N)}\,,
\end{equation}
again neglecting unimportant factors $2$. It is interesting to notice that, depending on how large the factor $(\epsilon\pi_0^2)/\lambda^2 N^2$ is, two different situations may be realized.
\begin{enumerate}
    \item When $(\epsilon\pi_0^2)/\lambda^2 N^2\gtrsim 1$, we have that $\sqrt{\epsilon/(\lambda N)}\gtrsim \pi_0^{-1}$, and so the condition \eqref{eqn:largensmallfluctx} in turns implies that $\vert\chi_0\vert \gg \pi_0^{-1}$. This condition, as shown in the above subsection, in turns implies that $\chi_0$ can be interpreted as the expectation value of the massless scalar field operator. 
    \item When instead $(\epsilon\pi_0^2)/\lambda^2 N^2\ll 1$, it may be that 
    \begin{equation*}
        \sqrt{\epsilon/(\lambda N)}\ll\vert\chi_0\vert \ll \pi_0^{-1}\,,
    \end{equation*}
    thus leading to a small relative variance of the massless scalar field operator but to the impossibility of identifying $\chi_0$ as a relational parameter after all.
\end{enumerate}
\end{description}
\paragraph*{Second case.} The arguments exposed in the first case about the variances of all the operators besides the volume operator (which after all just made reference to $N$, which is still $\gg 1$), are still valid. On the other hand, the inequalities used for the volume operator become inadequate in this case. Still, it is clear that we have
\begin{equation*}
    \rho_{j,\min}^2f_\Lambda\le \sigma^2_V\le \rho_{j,\min}^{-4}f_\Lambda\,,
\end{equation*}
where $\rho_{j,\min}^2\equiv \min_{j\in J}\rho_j^2$ and 
\begin{equation*}
    f_\Lambda\equiv\frac{\sum_{j=0}^\Lambda V_j^2}{\left(\sum_{j=0}^\Lambda V_j\right)^2}=\frac{\sum_{j=0}^\Lambda j^{3}}{\left(\sum_{j=0}^\Lambda j^{3/2}\right)^2}\simeq 3.56 \Lambda^{-3.49}\,,
\end{equation*}
where the last approximate equality has been obtained by an explicit fit. Notice that, since $N\simeq \sum_j\rho_j^2\gg 1$, we must have
\begin{equation*}
    \Lambda\rho_{j,\min}\le N\le \Lambda\,,
\end{equation*}
which are however not particularly helpful in extracting tighter bounds. As a conclusion, by defining $a=3.56$ and $b=3.49$, we see that, when
\begin{equation*}
    a\rho_{j,\min}^2/\Lambda\gtrsim 1\,,
\end{equation*}
fluctuations on the volume operator are certainly large, while when
\begin{equation*}
    a\rho_{j,\min}^{-4}\Lambda^{-b}\ll 1\,,
\end{equation*}
the relative volume variance is certainly negligible. This is of course the case when all the $\rho_j^2$s are of order one, since $\Lambda\gg 1$. 
\subsubsection{Number of GFT quanta of order of or smaller than one}\label{subsub:smalln}
When the number of GFT quanta is $N\lesssim 1$, the situation is far more complicated. In fact, not only all the relative variances computed so far can be large, but in this case one does not expect a hierarchy of moments of quantum operators, so that considering only relative variances in order to asses the possible quantum effects is no more enough. Furthermore, one expects also that in this regime a hydrodynamic approxiamtion cannot capture anymore the quantum dynamics of the fundamental \virgolette{atoms of spacetime}, which can only be consistently determined by solving all the Schwinger-Dyson equations of the theory and which is however pre-geometric and not in principle relational (as we intend it from the classical perspective). 

In such a case, therefore, not only we expect the CPSs not to define a notion of relational dynamics, but we expect averaged results not to capture all the relevant physics of the system. Hence, we will leave the study of this specific regime to some future work. 

\section{Effective relational dynamics:\\ the impact of quantum effects}\label{sec:quantumflucerd}
Let us now recapitulate our results and draw some conclusions from them.

In order for the cosmological CPS construction to fit in an effective relational framework, a certain number of conditions, proposed in \cite{Marchetti:2020umh} and reviewed in Subsection \ref{subsub:generalerd}, should be satisfied. Here, we summarize in which regimes they are satisfied, ensuring the reliability of the cosmological evolution obtained in \cite{Marchetti:2020umh}, with its classical Friedmann-like late times dynamics and singularity resolution into a bounce.

As mentioned in Subsection \ref{subsub:generalerd}, variances are not in general enough to characterize the properties of operators in a fully quantum regime (see also Subsection \ref{subsub:smalln}), except when there is a clear hierarchy among operator moments, with the higher ones being suppressed by higher powers of the number of quanta. If we try to quantify quantum fluctuations in terms of relative variances, as we will mostly do here, we must be careful not to assume that certain features characterizing the behavior of relative variances are true also for higher moments, since in certain regimes, variances may be indeed small but higher moments become relevant.  Still, as we have mentioned, we do expect that there exists a regime in which the aforementioned hierarchy among moments is indeed present: it is the case in which the number of GFT quanta is large. 

While in mesoscopic regimes it is not possible to determine under which conditions the hydrodynamic and the effective relational approximations are satisfied only by studying relative variances, large variances can however be taken as a clear evidence that one, or possibly both the above approximations are not adequate.

\subsection{Quantum effects in the effective relational CPS dynamics}\label{subsec:quantumeffectsrelational}
First, therefore, let us discuss the form that the conditions in \ref{subsub:generalerd} take in the CPS cosmology framework, focusing on the volume operator.
Then equation \eqref{eqn:averageddynamics} is satisfied by the CPS construction \cite{Marchetti:2020umh} provided that
\begin{enumerate}
    \item\label{item:necessary1} The expectation value of the (intrinsic) massless scalar field operator is $\chi_0$. 
\end{enumerate}
We have already mentioned that in general this is not exactly the case, essentially because we can not take the limit $\epsilon\to 0$ in order to avoid divergences in quantum fluctuations of the massless scalar field momentum. Hence, this issue is a consequence of the quantum properties of the chosen relational clock.

Also, in order to interpret the evolution generated by $\hat{H}$ as a truly relational one, we want its moments to coincide with those of $\hat{\Pi}$. Imposing this condition as an exact relation for the first moment and for any values of $\chi_0$ requires $\sum_j Q_j=0$, while this is not formally required in a large $N$ regime where the condition is satisfied approximately. 

A similar situation happens for the relative variance. Indeed, again in the large $N$ regime, $\delta^2_\Pi= \delta^2_H= N^{-1}$ to any degree of accuracy required\footnote{While a formal proof would be needed that similar results extend to even higher moments, it seems likely that this is the case.}.

On the other hand, let us notice that imposing the equality between \eqref{eqn:variancestrivial} and \eqref{eqn:finalsigmapi} for smaller $N$, and so for mesoscopic intermediate regimes, would impose another constraint on the initial conditions, requiring that all the $\alpha_j$s with $j\in P$ are zero. In turns, this implies that at least one of the $\alpha_j$s with $j\in J$ must be negative, in order not to have only trivial solutions. This means, from equation \eqref{eqn:approximatenv}, that the expectation value of the volume never vanishes, which might have important consequences for the volume evolution. 

Next, according to the general discussion in \cite{Marchetti:2020umh} and in Subsection \ref{subsub:generalerd}, one has to be sure that the clock variable is not \virgolette{too quantum}, which, in our framework can be phrased as the requirement that
\begin{enumerate}
\setcounter{enumi}{1}
    \item\label{item:necessary2} Relative quantum fluctuations of $\hat{\chi}$ must be much smaller than one.
\end{enumerate}
As usual, we can get some information about the behavior of clock quantum fluctuations from the form of $\delta_\chi^2$ obtained above. From the explicit expression in equation \eqref{eqn:variancechi}, we notice that besides the general behavior $\sim N^{-1}$, there is an additional irreducible contribution to quantum fluctuations parametrized by $\epsilon$. From the computations in the above subsection we conclude that in this case the smallness of the relative variance is dictated by a non-trivial interplay between $\epsilon$, $\chi_0$ and $N$, contrarily to what happens for the other observables. This might make it more difficult to try to extrapolate general features of even higher moments in a mesoscopic regime from those that we observe from the relative variance. 

Conditions \ref{item:necessary1} and \ref{item:necessary2} are the two necessary conditions that need to be satisfied in order to qualify the framework constructed so far as a truly relational one. 

\

Further, the evolution of the expectation value of the volume operator is a good enough characterization of the universe evolution (in the homogeneous and isotropic context) if
\begin{enumerate}
\setcounter{enumi}{2}
    \item\label{item:volume1} Quantum fluctuations (encoded in moments higher than the first one) of the volume operator are negligible.
\end{enumerate}
Also for this operator, as in the massless scalar field case, the existence of a hierarchy of moments is in general far from being trivial, since the relative variance is already strictly dependent on the possible spin cut-off scale.

However, even when satisfying condition \ref{item:volume1}, the resulting system might be highly non-classical, depending on the value of quantum fluctuations for the remaining operators $\hat{N}$, $\hat{\Pi}$ and $\hat{H}$. A necessary condition for a classicalization of the system to happen is therefore that
\begin{enumerate}
\setcounter{enumi}{3}
    \item\label{item:alloperators} Quantum fluctuations (encoded in moments higher than the first one) of all the relevant operators ($\hat{N}$, $\hat{\chi}$, $\hat{\Pi}$, $\hat{H}$, $\hat{V}$) are negligible.
\end{enumerate}
Therefore, in order for a classical relational regime to be realized at late enough times in the CPS framework, both conditions \ref{item:necessary1} (together with the identification of the moments of $\hat{H}$ and $\hat{\Pi}$) and \ref{item:alloperators} should be satisfied. In particular, large variances of any of these operators actually signal a breakdown of the hydrodynamic approximation underlying equation \eqref{eqn:simplestschwinger}.
\subsection{Effective relational volume dynamics with CPSs}\label{subsec:effectivevolumecps}
In light of the above conditions it is interesting to examine the relational evolution of the average of the volume operator, since, in GFT cosmology, it is at this level that the comparison with the Friedmann dynamics is usually performed. We will review this below, in Subsection \ref{subsub:generalpropertiesvolume}, emphasizing two main regimes of its evolution: a possible bounce and a Friedmann-like late evolution. In Subsection \ref{subsub:volumedynamicsquantumfluct}, instead, we will draw some general conclusions on the relationality and classicality of these two phases in light of the results obtained in the previous sections.
\subsubsection{General properties of the volume evolution}\label{subsub:generalpropertiesvolume}
Let us start from the general expression \eqref{eqn:completevolume}
\begin{equation*}
    V(\chi_0)=\sum_jV_j    \rho_j^2(\chi_0)\left[1+\sum_{n=1}^{\infty}\frac{[\rho_j^2]^{(2n)}(\chi_0))}{\rho_j^2(\chi_0)}\frac{\epsilon^{n}}{4^{n} n!}\right].
\end{equation*}
We see that $V(\chi_0)$ is always positive and never reaches zero, at least as long as one of the $\beta_{j}$s (equivalently, one of the $Q_{j}$) is different from zero. Also, from the above expression, we see that
\begin{subequations}
\begin{align}
V'&=\sum_{j}V_{j}C_{j}\mu_{j}\sinh\left(2\mu_{j}(\chi_{0}-\chi_{0,j})\right), \\ V''&=2\sum_{j}V_{j}C_{j}\mu_{j}^2\cosh\left(2\mu_{j}(\chi_{0}-\chi_{0,j})\right),
\end{align}
\end{subequations}
where
\begin{equation*}
C_{j}=\vert\alpha_j\vert\sqrt{1+r_j}\left(1+\sum_{n=1}^\infty\frac{(\epsilon\mu_j^2)^n}{n!}\right).
\end{equation*}
Since $V'\to\pm \infty$ when $\chi_{0}\to\pm \infty$, it has to cross zero\footnote{Here we are assuming $\mu_{j}>0$; if $\mu_{j}<0$ the limits are opposite, but the result is the same.}, and since $V''>0$ always, we see that $V'$ is monotone, so it has only one zero. This means that there is only one turning point. Were this evolution truly relational, the scenario would be that of a bouncing universe, increasing monotonically as the bounce happens, lately behaving as a Friedmann universe, as already suggested in \cite{Oriti:2016qtz} . Let us discuss this two features in more detail.
\begin{description}
\item[Bounce] 
The bounce happens at a relational time $\overline{\chi_{0}}$ which is $\min_{j\in J}\chi_{0,j}\le \bar{\chi}_{0}\le \max_{j\in J}\chi_{0,j}$. Indeed, $V'(\chi_{0})\vert_{\max_{j\in J}\chi_{0,j}}>0$, while $V'(\chi_{0})\vert_{\min_{j\in J}\chi_{0,j}}<0$. By continuity and monotonicity, the value of the bounce must be included among these two points. In addition, we notice from equation \eqref{eqn:approximatenv} and \eqref{eqn:generalsolution} that when at least one of the $r_j$s is different from zero, or at least one of the $\alpha_j$s is strictly negative, the volume never reaches zero. So, in these cases, the classical singularity is resolved into a bounce with non-zero volume\footnote{See \cite{Marchetti:2020umh, Oriti:2016qtz} for a comparison with LQG effective bouncing dynamics and \cite{Battefeld:2014uga} for a review of bouncing models.}.
\item[Friedmann regime] When $2\vert\mu_{j} (\chi_{0}-\chi_{0,j})\vert \gg 1$ (for each $j$), the hyperbolic functions can be approximated as simple exponentials. In that case, then, assuming $\mu_{j}$ is independent of $j$, (or that at least it is mildly dependent on it) one obtains 
\begin{equation}
\left(\frac{V'}{3V}\right)^2=\left(\frac{2}{3}\mu_{j}\right)^2\,,\qquad V''/V=4\mu_{j}^2\,,
\end{equation}
which are indeed the flat space ($k=0$) Friedmann equations, upon imposing that $\mu_{j}^2= \mu^2=3\pi \tilde{G}$ ($\tilde{G}$ being the dimensionless gravitational constant) \cite{Marchetti:2020umh} (see Appendix \ref{app:classical}, equations \eqref{eqn:friedmannrelational}). Notice that the same equations can be obtained also when one of the $j$s is dominating the volume evolution, and its corresponding $\mu_{j}$ is then taken to be proportional to the (effective) Newton constant \cite{Wilson-Ewing:2018mrp}.
\end{description}
\subsubsection{Effective relational volume dynamics with CPS:\\ the impact of quantum effects}\label{subsub:volumedynamicsquantumfluct}
Let us now characterize better these two phases of the evolution of the average volume in terms of their relationality and quantumness.
\paragraph*{Friedmann dynamics and classical regime.}
The Friedmann regime is selected by the condition $\vert x_j\vert \gg 1$ for each $j$. Notice that this condition does not necessarily imply that $\rho_j^2\gg 1$, but it does imply that
\begin{equation*}
    \sum_{j\in P}\alpha_j\ll \sum_{j\in P}\rho_j^2\lesssim N\,.
\end{equation*}
As the parameter $x_j$ grows, eventually it will be far enough from each single $x_j^o$ to make the factor $\vert \Delta X\vert/N\ll 1$, and, eventually also making $N$ arbitrarily large. From these conditions we see that all the fluctuations on the relevant operators become negligible, and the parameter $\chi_0$ becomes the expectation value of the massless scalar field operator $\hat{\chi}$. Therefore, all the conditions from \ref{item:necessary1} to \ref{item:alloperators} (including the matching of all the moments of $\hat{H}$ and $\hat{\Pi}$) are satisfied. As a result, 
\begin{description}
\item[Statement $1$]
\emph{For the chosen approximation of the underlying GFT dynamics, a classical regime in which the volume evolution with respect to $\chi_0$ can be interpreted as a relational flat space Friedmann dynamics with respect to a massless scalar field clock is always realized, independently of the initial conditions.}
\end{description}
\begin{figure*}
    \centering
    \includegraphics[width=0.49\textwidth]{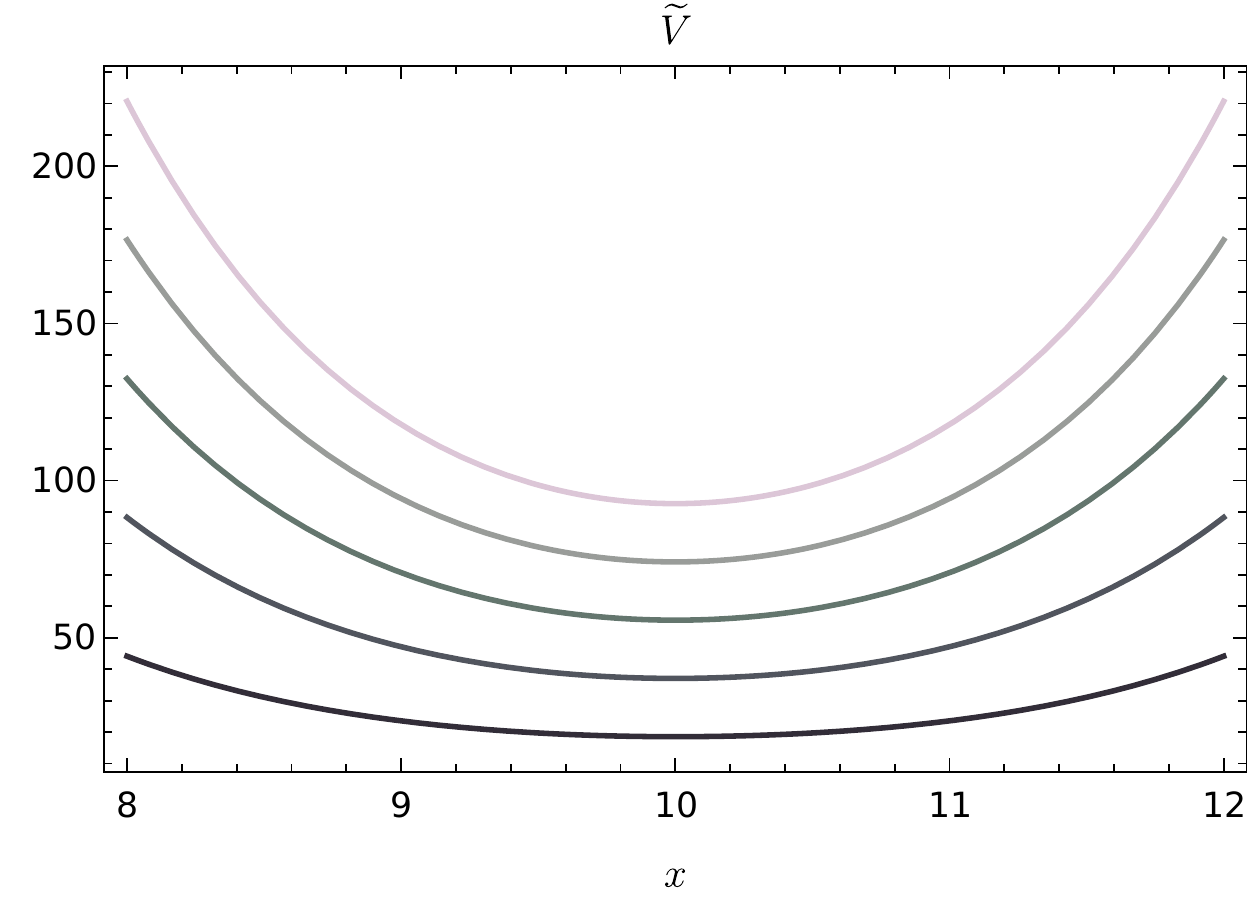}
    \includegraphics[width=0.49\textwidth]{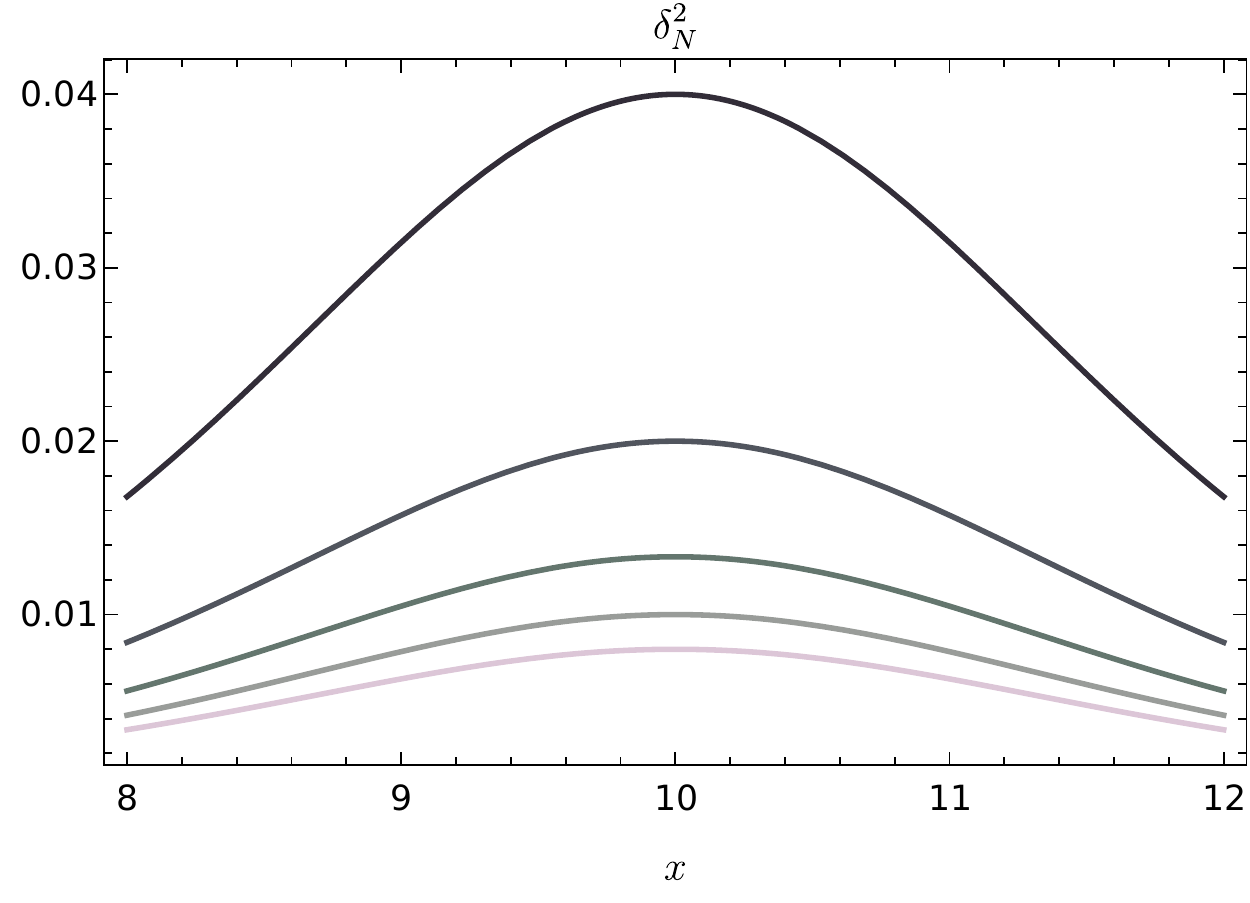}
    \caption{Plots of the dimensionless volume operator $\tilde{V}\equiv V/L_\text{Pl}$ and of the relative variance of the number operator $\delta_N^2$ as functions of $x=2\mu \chi_0$ in a two-spin scenario with $j_1=1/2$ and $j_2=1$. The plots are obtained with $\mu_{j_1}=\mu_{j_2}=\mu$, $r_{j_1}=r_{j_2}=0$, $x_{0,j_1}=x_{0,j_2}=2\mu\bar{\chi}_0=10$, and $\alpha_{j_1}=-10c$, $\alpha_{j_2}=-15c$, with $c$ varying from $1$ to $5$ in integer steps. Darker (lighter) lines correspond to smaller (higher) values of $c$.}
    \label{fig:varyingalpha}
\end{figure*}
Let us remark that the approximations involving the underlying GFT dynamics that we used to extract an effective mean field dynamics (see Subsection \ref{subsec:redwavefunctiondynamics}) may be very important for the validity of the above statement. 
For instance, among those, a crucial one was the approximation of negligible interactions\footnote{Let us also mention, however, that another important role is played by the assumption $\vert B_{j}/A_{j}\vert\ll \epsilon^{-1}$, see footnote \ref{footnote:detailsgft}.}. These, however, are supposed to become relevant as the average number of GFT quanta become very large, which is the asymptotic regime in which conditions $1$ and $4$ are expected to be satisfied. When interactions become important it is certainly possible that some of the above arguments do not hold anymore, but it is also possible that non-zero interactions do not modify substantially the conclusions above, but they change the effective matter content of the Friedmann system, possibly including now a dark sector (see e.g.\ \cite{Pithis:2016cxg,Pithis:2016wzf, Oriti:2021rvm}).

\paragraph*{Bounce.}
The situation concerning the bounce is much more complicated, essentially because it is not an asymptotic regime, and thus the value of initial conditions turns out to be important. It is less obvious whether the bounce can be interpreted as a relational dynamics result, or if the averaged evolution is overwhelmed by quantum fluctuations, thus making us question also the validity of the hydrodynamic approximation in \eqref{eqn:simplestschwinger}. 

In general, it might happen that both the conditions \ref{item:necessary1} and \ref{item:necessary2} are not satisfied. For instance, this is the case if the bounce happens at $\overline{\chi_0}=0$ with initial conditions such that $N(0)\lesssim 1$. Similarly, it might happen that only one of the two conditions above is satisfied. This is the case, for instance, of having a bounce $\overline{\chi_0}$ such that
\begin{equation*}
        \sqrt{\epsilon/(\lambda N)}\ll\vert\overline{\chi_0}\vert \ll \pi_0^{-1}\,,
    \end{equation*}
with arbitrarily large values of $N(\overline{\chi_0})$, so that essentially the bounce happens already in a \lq large volume\rq $\,$regime. In this case quantum fluctuations of all the relevant operators are negligible but the interpretation of $\overline{\chi_0}$ as expectation value of the massles scalar field operator is not allowed. Or, the other way around, it might be that indeed $\overline{\chi_0}\gg \pi_0^{-1}$, thus allowing to interpret $\overline{\chi_0}$ as expectation value of $\hat{\chi}$ but $N(\overline{\chi_0})\lesssim 1$, making fluctuations possibly very large for all the relevant operators.

On the other hand, there are regimes in which a bounce can satisfy all the conditions from \ref{item:necessary1} to \ref{item:alloperators}. For instance, let us consider the case in which all the $\chi_{0,j}\equiv \overline{\chi_0}$, which therefore marks the bounce. Also, let us assume that $\rho_j^2(\overline{\chi_0})\gg 1$ for each $j\in J$, so that $N(\overline{\chi_0})\gg 1$ too. Let us fix $\overline{\chi_0}>0$, in particular with $\overline{\chi_0}\gg\sqrt{\epsilon/N(\overline{\chi_0})}$. Lastly, let us also assume that $r_j\gtrsim 1$ for each $j\in P$. Then we know that $\Delta X/N$ is negligible for each $\chi_0\ge \overline{\chi_0}$, and that relative variances are negligibly small. More precisely, relative variances of $\sigma^2_N$, $\sigma^2_V$ and $\sigma^2_H$ are small because of $\rho_j^2(\overline{\chi_0})\gg 1$,  $\sigma^2_\Pi$ is small because of $N(\overline{\chi_0})\gg 1$ and $r_{j\in P}\gtrsim 1$, while $\sigma^2_\chi$ is small because of $N(\overline{\chi_0})\gg 1$ and $\overline{\chi_0}\gg \sqrt{\epsilon/N(\overline{\chi_0})}$. If we could rest assured that all moments higher than the second one are negligible as well and that the effective equality between $\hat{\Pi}$ and $\hat{H}$ is guaranteed, then we could conclude that the bouncing scenario would be not only reliable and truly relational, but that it could also admit an effective classical description (in terms of some modified gravity theory, with an interesting possibility being mimetic gravity \cite{deCesare:2018cts}). 
Notice also, that under the above conditions, the dynamics is indeed relational from the point $\overline{\chi_0}$ on. In practice, therefore, when these conditions are realized, one could follow the volume evolution from the bouncing point to the Friedmann regime and on towards infinite values of $\chi_0 $.

\begin{figure}
    \centering
    \includegraphics[width=\linewidth]{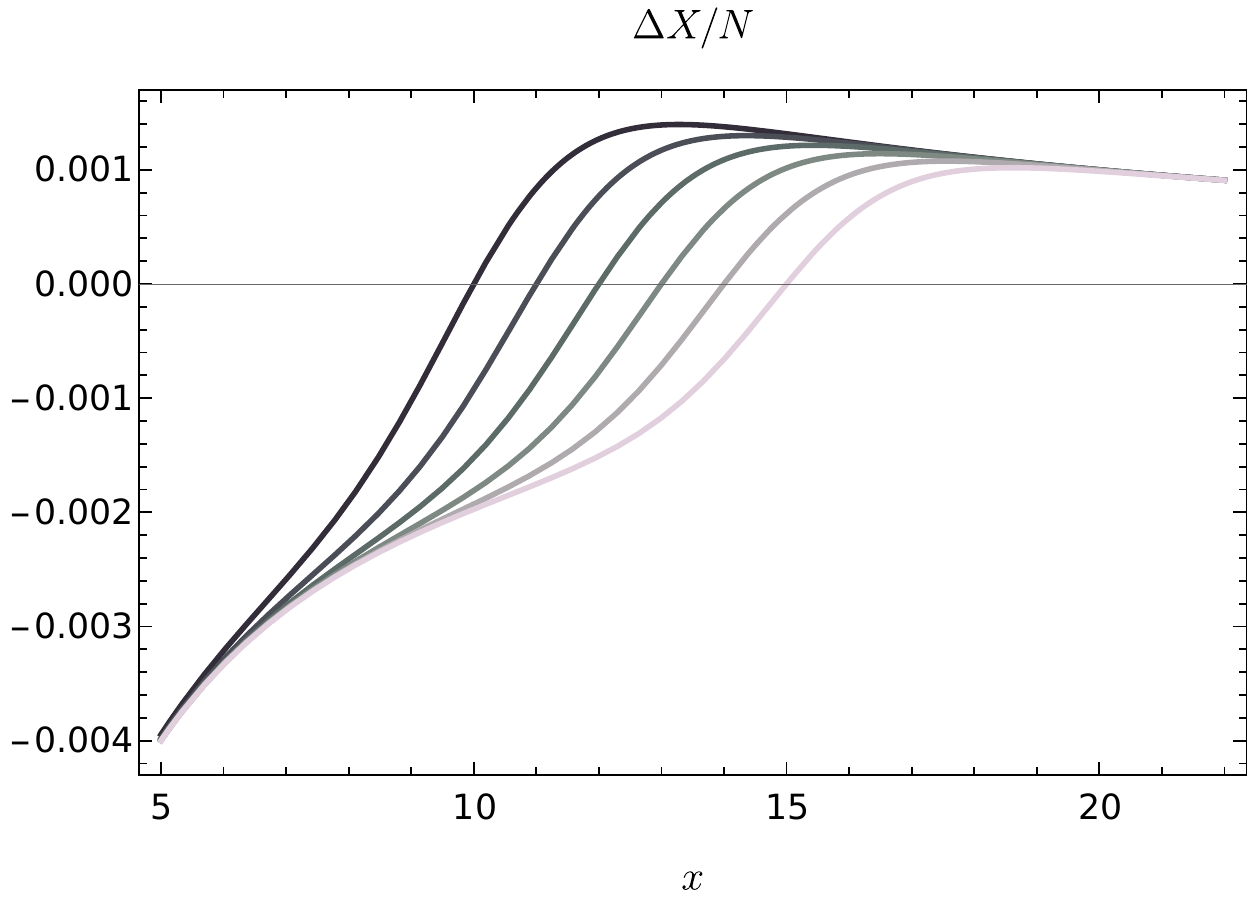}
    \caption{Plot of $\Delta X/N$ as a function of $x$ with  $\mu_{j_1}=\mu_{j_2}=\mu$, $r_{j_1}=r_{j_2}=0$, $\alpha_{j_1}=-10$, $\alpha_{j_2}=-15$ and $\epsilon\mu^2=10^{-2}$, but with different values of the bounce, given by $x_{j_1}=x_{j_2}=(1+0.1c)10$ for $c$ going from $0$ to $5$ in integer steps. Darker (lighter) lines correspond to smaller (higher) values of $c$.}
    \label{fig:deltax}
\end{figure}
\begin{figure}
    \centering
    \includegraphics[width=\linewidth]{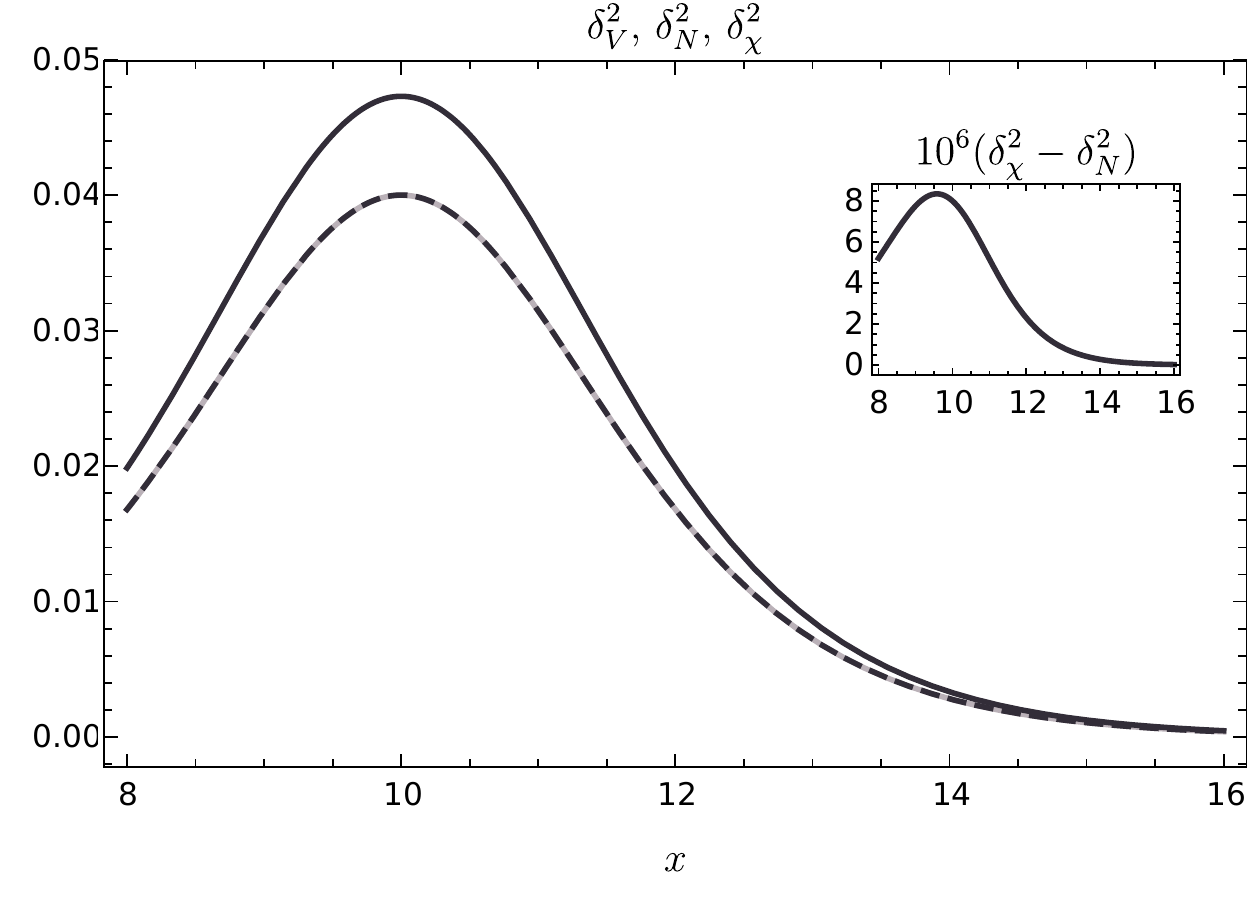}
    \caption{Plot of $\delta_V^2$ (dark solid line), $\delta_\chi^2$ (dark dashed line) and $\delta_N^2$ (light solid line) as functions of $x\equiv 2\mu \chi_0$ around the bounce $\bar{x}=10$ for $\mu_{j_1}=\mu_{j_2}=\mu$, $r_{j_1}=r_{j_2}=0$, $x_{0,j_1}=x_{0,j_2}=2\mu\bar{\chi}_0=10$, and $\alpha_{j_1}=-10$, $\alpha_{j_2}=-15$. In the inset plot, instead, is represented $10^6(\delta_\chi^2-\delta_N^2)$ for the same choice of the relevant parameters. }
    \label{fig:allfluctuations}
\end{figure}

The relevant quantities for a simplified two-spin scenario like the one described above with $j_1=1/2$, $j_2=1$, $\mu_{j_1}=\mu_{j_2}=\mu$, $\epsilon\mu^2\ll 1$, $r_{j_1}=r_{j_2}=0$, and $\alpha_j<0$, $\vert\alpha_j\vert\gg 1$ for $j\in\{j_1,j_2\}$ are plotted in Figures\footnote{Indeed, under these assumptions expectation values and variances of $\hat{\Pi}$ and $\hat{H}$ are determined by $N$.}  \ref{fig:varyingalpha}, \ref{fig:deltax} and \ref{fig:allfluctuations} as functions of $x\equiv 2\mu \chi_0$. The value of the bounce $\chi_{0,j_1}=\chi_{0,j_2}=\bar{\chi}_0$ is taken to be far enough from $\chi_0=0$, so to avoid any unphysical singularity in the quantities represented. In Figure \ref{fig:varyingalpha}, we vary the values of $\alpha_j$ from lower to higher values (darker to lighter colors in the plots). In the left panel, we represent the dimensionless volume operator $\tilde{V}=V/L_\text{Pl}^3$ \cite{Oriti:2016qtz} (i.e., such that $\tilde{V}_j\sim j^{3/2}$), while on the right panel, we represent the variance of the number operator. As the values of $\alpha_j$ are increased, the minimum value of the averaged volume becomes larger, while $\delta^2_N$ becomes less and less important at the bounce. This behavior is shared also by $\delta_\chi^2$ and $\delta_V^2$, since $\delta^2_N$ sets the scaling of the relative variances of all the operators. Indeed, as we can see from Figure \ref{fig:allfluctuations}, they are of the same order of magnitude. Actually, one notices that fluctuations in $\hat{\chi}$ and in $\hat{N}$ (dashed dark line and lighter solid line respectively) are very close to each other, with differences only of order $10^{-5}$-$10^{-6}$ in the range plotted. This is due to the smallness of the quantity $\Delta X/N$, which is plotted in Figure \ref{fig:deltax} for increasing values of $\bar{\chi}_0$, as we see from equation \eqref{eqn:variancechi}. From both Figures \ref{fig:deltax} and \ref{fig:allfluctuations}, we notice that all the variances and $\Delta X/N$ go to zero at large positive $x$ (where the Friedmann regime is expected to kick in). 
In any case, we should remark that,  since we currently have little control on moments higher than the second ones, one can take the above example only as an indication of the existence of the singularity resolution into a bounce.

In general, therefore, we can draw the following conclusion:
\begin{description}
\item[Statement $2$] \emph{The bouncing scenario is not a universal feature of the model, meaning that it is not realized under arbitrary choices of the initial conditions. However, if (i) there exists at least one $\alpha_j<0$ or at least one $r_j\neq 0$, (ii) all the quantities in equations \eqref{eqn:deltaxnsimp}, \eqref{eqn:variancechi} and \eqref{eqn:variancevolumefinal} are much less than one when the averaged volume attains its (non-zero) minimum, and (iii) all the higher moments of the volume and massless scalar field are negligible, the initial singularity is indeed resolved into a bounce\footnote{Notice that the requirements (ii) and (iii) correspond to conditions from $1$ to $3$ being satisfied. The first two of them qualify the framework as relational, while the third one guarantees that the expectation value of the volume operator captures in a satisfactory way the relational evolution of the homogeneous and isotropic geometry.}.}
\end{description}
We remark again that this lack of universality is due to the possible role of quantum fluctuations, in particular higher moments, which may make the relational evolution unreliable, while the bouncing dynamics of the average universe volume is in fact general (but not necessarily with a non-zero minimum value). In other words, whether or not the dynamics of the volume is relational and entirely captured by the lowest moment strongly depends on the initial conditions.
\subsubsection{An example: single spin condensate}\label{subsub:singlespin}
As an explicit and fairly simple (though possibly very physically relevant \cite{Gielen:2016uft}) example of the arguments exposed above, let us consider the case in which only one spin among those in $J$ is excited, say $j_o$, so that all the sums characterizing the collective operators above are not present anymore. For instance, we now have
\begin{equation}\label{eqn:nvsinglespin}
    N(\chi_0)\simeq\rho_{j_o}^2(\chi_0)\,,\qquad V(\chi_0)\simeq V_{j_o}\rho_{j_o}^2(\chi_0)\,,
\end{equation}
where
\begin{equation}\label{eqn:rhojsinglespin}
    \rho_{j_o}^2(\chi_0)=\frac{\vert\alpha_{j_o}\vert}{2}\left(-\text{sgn}(\alpha_{j_o})+\cosh x_{j_o}\right)\,,
\end{equation}
where we have imposed the condition $\sum_jQ_j=0$, i.e., $Q_{j_o}=0$, or $\beta_{j_o}=0$, since we would like to have a relational framework even in intermediate regimes.

Let us study in detail under which conditions a resolution of the initial singularity into a bouncing universe, assuming that indeed quantum effects are effectively encoded into relative variances (so that we can neglect the impact on the system of moments of relevant operators higher than the second one). From equations \eqref{eqn:nvsinglespin} and \eqref{eqn:rhojsinglespin}, we deduce that a bounce with a non-zero value of the (average) volume happens only when $\alpha_{j_o}<0$. We also recall that in this case one has an equality between the second moments of the Hamiltonian and the momentum operator. So, in the following, we will specialize to this case. The situation in this case simplifies considerably: for instance, we have
\begin{equation}\label{eqn:variancesssc}
    \delta^2_N=\delta^2_V=\sigma^2_H=\delta^2_\Pi=N^{-1}
\end{equation}

Before proceeding with further considerations, it is interesting to remark that the single spin case mirrors the situation appearing in Loop Quantum Cosmology (LQC) \cite{Bojowald:2008zzb,Ashtekar:2011ni}, where one considers a LQG fundamental state corresponding to a graph constructed out of a large number of nodes and links with the latter being associated all to the same spin. This similarity can be also observed in fluctuations. Indeed, from the above equation we see that in this case the quantity governing quantum fluctuations is exactly the average number of particles, with variances suppressed as $N^{-1}$ for large $N$. In LQC, the quantity setting the scale of quantum fluctuations is $V_0$ \cite{Rovelli:2013zaa}, the coordinate volume of the fiducial homogenous patch under consideration. In a graph interpretation of the LQC framework, $V_0=N\ell_0$, with $\ell_0$ being a fundamental coordinate length, adding another interesting \virgolette{phenomenological} connection besides those already presented in \cite{Gielen:2014uga,Oriti:2016qtz,Marchetti:2020umh} between these two approaches.

Going back to equation \eqref{eqn:variancesssc}, we see that, in order for the bounce to have any hope of being classical, we also need to require $\vert\alpha_{j_o}\vert \gg 1$. For the moment, therefore, the two conditions that we have imposed on $\alpha_{j_o}$ are
\begin{equation}\label{eqn:initialconditionssinglespin}
   \alpha_{j_o}<0\,,\qquad \vert\alpha_{j_o}\vert\gg 1\,.
\end{equation}
What is left to check are the values of $\Delta X/N$ and $\sigma^2_\chi$, which are required to be small in order to have a meaningful relational dynamics.
\begin{description}
\item[$\Delta X/N$] About $\Delta X/N$, the computation is straightforward: we have
\begin{align*}
    \frac{\vert\Delta X\vert}{N}&=\frac{\vert(\rho_{j_o}^2)'(\chi_0)\vert}{\vert\chi_0\vert}\frac{1}{\rho_{j_o}^2(\chi_0)}\frac{\epsilon}{2}\\&=\frac{\vert\sinh x_{j_o}\vert}{\vert x_{j_o}+x_{j_o}^o\vert}\frac{1}{1+\cosh x_{j_o}}\epsilon\mu_{j_o}^2\,.
\end{align*}
So, we conclude that for each $x_{j_o}^o\ge 0$ (i.e., for each $\chi_{0,j_o}\ge 0$) the above quantity is always $\ll 1$ for each $\chi_0\ge 0$. 
\item[$\delta^2_\chi$] About the relative variance of the massless scalar field operator, assuming $\chi_{0,j_o}\ge 0$, we have
\begin{equation*}
    \sigma^2_\chi=N^{-1}+\frac{\epsilon}{2N\chi_0^2}\,.
\end{equation*}
Since the first term is always much smaller than $1$ under our assumptions, the relative variance of the massless scalar field operator is negligible as long as $\epsilon/(2N\chi_0^2)$ is negligible as well. This is satisfied for each $\chi_0\ge \chi_{0,j_o}$ as long as $(\chi_{0,j_o})^2\gg\epsilon/(2\vert\alpha_j\vert)$.
\end{description}
Notice that the assumption of $\chi_{0,j_o}\ge 0$ ($\chi_{0,j_o}\le 0$) is necessary if one wants to have a relational picture extending from today to the bounce among positive (negative) values of the massless scalar field. Indeed, if the bounce had happened at, say $\chi_{0,j_o}<0$ (today being at positive values of the massless scalar field), we should have crossed the point $\chi_0=0$, which is however a point in which relative quantum fluctuations formally diverge and the clock may become not classical anymore. In conclusion, by further assuming that
\begin{equation}\label{eqn:conditionsclocksinglespin}
    \chi_{0,j_o}\ge 0\,,\qquad (\chi_{0,j_o})^2\gg\epsilon/(2\vert\alpha_{j_o}\vert)\,,
\end{equation}
the singularity is indeed replaced by a bounce (again assuming $\chi_0^{\text{today}}>0$). Notice that the second inequality above does not impose a very strict constraint on $\chi_{0,j_o}$, since by construction of the CPSs $\epsilon$ is assumed to be a very small quantity.

To sum up, a classical bounce that can be understood within the effective relational framework discussed above and in \cite{Marchetti:2020umh}, can be obtained in this single spin case for instance by requiring that
\begin{enumerate}
    \item $Q_{j_o}=0$, guaranteeing equality between the expectation value of $\hat{\Pi}$ and $\hat{H}$;
    \item conditions \eqref{eqn:initialconditionssinglespin} are satisfied, the first of which guarantees that the expectation value of the volume operator reaches a non-zero minimum before bouncing, and the second of which guarantees small relative variances of the operators $\hat{N}$, $\hat{V}$, $\hat{H}$ and $\hat{\Pi}$;
    \item assuming that $\chi_0^{\text{today}}>0$, conditions \eqref{eqn:conditionsclocksinglespin} are satisfied\footnote{If $\chi_0^{\text{today}}<0$ the first condition in \eqref{eqn:conditionsclocksinglespin} would read $\chi_{0,j_o}\le 0$. }. The first of them guarantees that $\chi_0$ can be interpreted as the expectation value of the (intrinsic) scalar field operator, while the second one guarantees that its relative quantum fluctuations stay small during the whole Universe's evolution from the bounce until today.
\end{enumerate}

Under these assumptions, the relational time elapsed from the bounce would be 
\begin{align*}
    x^{\text{today}}_{j_o}&\simeq\log\left[\frac{V_{\text{today}}}{V_{j_o}}\frac{2}{\vert\alpha_{j_o}\vert}-1\right]\\
    &\simeq\log\left[\frac{V_{\text{today}}}{V_{j_o}}\frac{2}{\vert\alpha_{j_o}\vert}\right]=\log\frac{V_{\text{today}}}{V_{j_o}}-\log\frac{\vert\alpha_{j_o}\vert}{2}\,,
\end{align*}
where we have assumed the term $-1$ to be negligible with respect to the first contribution. If we further assume that the right-hand-side of the last equality is dominated by the first term, we get
\begin{equation}
    x^{\text{today}}_{j_o}\simeq \log\frac{ V_{\text{today}}}{V_{j_o}}\sim 252-\frac{3}{2}\log j_o\,,
\end{equation}
where the last line is just the result of a crude estimate obtained from $V_{\text{today}}\sim H_0^{-3}\simeq (9.25 h\times 10^{25}\text{ m})^3$ , with $h\simeq 0.71$ and $V_{j_o}\simeq (L_P)^3 j_o^{3/2}$.
\nocite{*}

\section{Conclusions}
We have analysed the size and evolution of quantum fluctuations of cosmologically relevant geometric observables (in the homogeneous and isotropic case), in the context of the effective relational cosmological dynamics of quantum geometric GFT models of quantum gravity. We considered first of all the fluctuations of the matter clock observables, to test the validity of the relational evolution picture itself. Next, we studied quantum fluctuations of the universe volume and of other operators characterizing its evolution, like the number operator for the fundamental GFT quanta, the effective Hamiltonian and the scalar field momentum (which is expected to contribute to the matter density). In particular, we focused on the late (clock) time regime (see Statement $1$, Subsection \ref{subsub:volumedynamicsquantumfluct}), where the dynamics of volume expectatation value is compatible with a flat FRW universe, and on the very early phase near the quantum bounce. We found that the relative quantum fluctuations of all observables are generically suppressed at late times, thus confirming the good classical relativistic limit of the effective QG dynamics.
Near the bounce, corresponding to a mesoscopic regime in which the average number of fundamental GFT quanta can not be arbitrarily large, the situation is much more delicate (see Statement $2$, Subsection \ref{subsub:volumedynamicsquantumfluct}). Depending on the specific choice of parameters in the fundamental dynamics and in the quantum condensate states, relational evolution as implemented by the CPSs strategy may remain consistent or become unreliable, due to fluctuations of the clock itself and to possible issues with \virgolette{synchronization} of the fundamental GFT quanta. Even when the relational evolution picture remains valid, quantum fluctuations of the geometric observables may become large, depending again on the precise values of the various parameters. When this happens, this could signal simply a highly quantum regime, but one that is still describable within the hydrodynamic approximation in which the effective cosmological dynamics has been obtained; or it could be interpreted as a signal of a breakdown of the same hydrodynamic approximation, calling for a more refined approximation of the underlying quantum gravity dynamics of the universe.

The analysis will have now to be extended to the case in which GFT interactions are not negligible. We expect such interactions to be most relevant at late clock times and largish universe volume (i.e. largish GFT condensate densities) \cite{Marchetti:2020umh,Oriti:2016qtz}, thus it is unclear whether they should be expected to modify much the behaviour of quantum fluctuations, since the are suppressed in the same regime. However, GFT interactions also modify the underlying dynamics of the volume itself, possibly causing a recollapsing phase \cite{deCesare:2016rsf}, thus they may as well enhance quantum fluctuations in such cases. Another important extension would be of course the inclusion of anisotropies \cite{deCesare:2017ynn}, but this is something we need to control much better already at the level of expectation values of geometric observables, in order to be confident about the resulting physical picture. Finally, quantum fluctuations should be considered in parallel with thermal fluctuations, which we can now compute as well using the recently developed thermofield double formalism for GFTs \cite{Kotecha:2018gof,Assanioussi:2019ouq,Assanioussi:2020hwf}.

Thus, much more work is called for. It is clear, however, that we now have a solid context to tackle cosmological physics from within full quantum gravity, also for what concerns quantum fluctuations. While we move towards the analysis of cosmological perturbations \cite{Gielen:2018xph,Gielen:2017eco} and the associated quantum gravity phenomenology, these results will help to control better the viability of the picture of the evolution universe we will 
 obtain.
\section*{Acknowledgements}
Financial support from the Deutsche Forschunggemeinschaft (DFG) is gratefully acknowledged. LM thanks the University of Pisa and the INFN (section of Pisa) for financial support, and the Ludwig-Maximilians-Universität Munich for the hospitality.

\appendix
\section{Explicit computation of relative variances}
\label{app:explicitfluctuations}
\paragraph*{Number operator.}
The prototypical example of the computations we will perform here is given by the number operator. The expectation value of the square of the momentum operator is
\begin{align*}
\braket{\hat{N}^2}_{\sigma;\chi_{0},\pi_{0}}&=\int\diff g_I\int\diff h_{I}\int\diff\chi\int\diff\chi'\\
&\times\!\!\braket{\hat{\varphi}^\dagger(g_I,\chi)\hat{\varphi}(g_I,\chi)\hat{\varphi}^\dagger(h_I,\chi')\hat{\varphi}(h_I,\chi')}_{\sigma;\chi_{0},\pi_{0}}\!.
\end{align*}
By putting the operators in the brackets in normal ordering one gets
\begin{equation*}
\braket{\hat{N}^2}_{\sigma;\chi_{0},\pi_{0}}= N^2(\chi_{0})+N(\chi_{0})\,,
\end{equation*}
so that the relative variance is
\begin{equation*}
\delta^2_{N}= N^{-1}(\chi_{0})\,.
\end{equation*}
\paragraph*{Volume operator.}
Similar arguments hold for the volume operator, so that one obtains
\begin{equation*}
    \delta^2_V=\frac{\sum_jV_j^2\rho_j^2\left[1+\sum_{n=1}^{\infty}\frac{[\rho_j^2]^{(2n)}(\chi_0))}{\rho_j^2(\chi_0)}\frac{\epsilon^{n}}{4^{n} (n)!}\right]}{\left\{\sum_jV_j\rho_j^2\left[1+\sum_{n=1}^{\infty}\frac{[\rho_j^2]^{(2n)}(\chi_0))}{\rho_j^2(\chi_0)}\frac{\epsilon^n}{4^{n} (n)!}\right]\right\}^2}\,.
\end{equation*}
\paragraph*{Hamiltonian operator.}
Moving to the relative variance of the Hamiltonian operator, we notice that, with our definition of $\hat{\bar{H}}$, we have
\begin{align*}
&\braket{\hat{H}^\dagger \hat{H}}_{\sigma_{\epsilon};\chi_{0},\pi_{0}}-\braket{\hat{H}^\dagger}_{\sigma_{\epsilon};\chi_{0},\pi_{0}}\braket{\hat{H}}_{\sigma_{\epsilon};\chi_{0},\pi_{0}}\\\quad&= \braket{\hat{\bar{H}}^\dagger \hat{\bar{H}}}_{\sigma_{\epsilon};\chi_{0},\pi_{0}}-\braket{\hat{\bar{H}}^\dagger}_{\sigma_{\epsilon};\chi_{0},\pi_{0}}\braket{\hat{H}}_{\sigma_{\epsilon};\chi_{0},\pi_{0}}\, ,
\end{align*}
and we can evaluate the relative variance of $\hat{H}$ on our CPSs by computing
\begin{widetext}
\begin{align*}
\braket{\hat{\bar{H}}^\dagger \hat{\bar{H}}}_{\sigma_{\epsilon};\chi_{0},\pi_{0}}&=\int\diff g_I\diff h_I\int\diff \chi\diff \chi'\,\bra{\sigma_{\epsilon};\chi_{0},\pi_{0}}\hat{\varphi}(h_I,\chi')\partial_{\chi'}\overline{\eta}_{\epsilon}(\chi'-\chi_{0},\pi_{0})\overline{\tilde{\sigma}}(h_I,\chi')\\
&\quad\quad\times\hat{\varphi}^\dagger(g_I,\chi)\partial_{\chi}\eta_{\epsilon}(\chi-\chi_{0},\pi_{0})\tilde{\sigma}(g_I,\chi)\ket{\sigma_{\epsilon};\chi_{0},\pi_{0}}\\
&=\braket{\hat{\bar{H}}^\dagger}_{\sigma_{\epsilon};\chi_{0},\pi_{0}}\braket{\hat{\bar{H}}}_{\sigma_{\epsilon};\chi_{0},\pi_{0}}+\int\diff g_I\int\diff\chi\,\partial_{\chi}\overline{\eta}_{\epsilon}(\chi-\chi_{0},\pi_{0})\partial_{\chi}\eta_{\epsilon}(\chi-\chi_{0},\pi_{0})\vert\tilde{\sigma}(g_I,\chi)\vert^2 \quad .
\end{align*}
\end{widetext}
The second term above gives 
\begin{align*}
&\int\diff g_I\int\diff\chi\left[\frac{(\chi-\chi_{0})^2}{\epsilon^2}+\pi_{0}^2\right]\vert\eta_{\epsilon}(\chi-\chi_{0},\pi_{0})\vert^2\vert\tilde{\sigma}(g_I,\chi)\vert^2\\
&\quad\simeq\pi_{0}^2 N(\chi_{0})\left(1+(2\epsilon\pi_{0}^2)^{-1}\right) \,.
\end{align*}
In conclusion we obtain
\begin{align}
\delta^2_{H}&\equiv \frac{\braket{H^\dagger H}_{\sigma_{\epsilon};\chi_{0},\pi_{0}}-\braket{\hat{H}^\dagger}_{\sigma_{\epsilon};\chi_{0},\pi_{0}}\braket{\hat{H}}_{\sigma_{\epsilon};\chi_{0},\pi_{0}}}{\braket{\hat{H}^\dagger}_{\sigma_{\epsilon};\chi_{0},\pi_{0}}\braket{\hat{H}}_{\sigma_{\epsilon};\chi_{0},\pi_{0}}}\nonumber\\&\simeq N^{-1}(\chi_{0})\left[1+\left(2\epsilon\pi_{0}^2\right)^{-1}\right]\simeq N^{-1}(\chi_0)\,.
\end{align}
\paragraph*{Momentum operator.}
Next, we discuss the variance of the momentum operator. The computation is slightly longer than the previous ones, but in the end one finds that
\begin{align*}
&\braket{\hat{\Pi}^2}_{\sigma;\chi_{0},\pi_{0}}=\braket{\hat{\Pi}}^2_{\sigma;\chi_{0},\pi_{0}}\\
&\quad+\sum_{j}\int\diff\chi\partial_{\chi}\overline{\sigma}_{\epsilon,j}(g_I,\chi;\chi_{0},\pi_{0})\partial_{\chi}\sigma_{\epsilon,j}(g_I,\chi;\chi_{0},\pi_{0})\,.
\end{align*}
After decomposing $\sigma_{\epsilon,j}\equiv \rho_{\epsilon,j}\exp[i\theta_\epsilon,j]$, with
\begin{equation*}
\rho_{\epsilon,j}\equiv \mathcal{N}_{\epsilon}e^{-(\chi-\chi_{0})^2/(2\epsilon)}\rho_{j}\,,\qquad \theta_{\epsilon,j}\equiv \theta_{j}+\pi_{0}(\chi-\chi_{0})\,,
\end{equation*}
the second line in the above equation becomes 
\begin{align*}
\sum_{j}\int\diff \chi&\biggl[\left(\partial_\chi\rho_{\epsilon,j}(\chi;\chi_0)\right)^2+\rho^2_{\epsilon,j}(\chi;\chi_0)(\partial_\chi\theta_{\epsilon,j}(\chi;\chi_0,\pi_{0}))^2\biggr] \, .
\end{align*}
The explicit evaluation of the derivatives give
\begin{align}
\delta \Pi^2_{\sigma_\epsilon;\chi_0,\pi_0}&\equiv \braket{\hat{\Pi}^2}_{\sigma_\epsilon;\chi_0,\pi_0}-\braket{\hat{\Pi}}_{\sigma_\epsilon;\chi_0,\pi_0}\nonumber\\
&=\sum_{j}\mathcal{N}_{\epsilon}^2\int\diff\chi\rho_{j}^2e^{-\frac{(\chi-\chi_{0})^2}{\epsilon}}\label{eqn:deltamomentum}\\
&\qquad\times \left[\left[\frac{\chi-\chi_{0}}{\epsilon}-\frac{1}{2}\frac{[\rho_{j}^2]'}{\rho_{j}^2}\right]^2\!\!+(\theta_{j}'+\pi_{0})^2\right].\nonumber
\end{align}
The evaluation of this integral can be quite complicated, so it is useful to start by simplifying the above expression. First, we notice that the last term can be written approximately as
\begin{equation}\label{eqn:approxpi}
(\theta_{j}'+\pi_{0})^2\simeq \pi_{0}^2+2\pi_{0}Q_{j}/\rho_{j}^2+Q_{j}^2/\rho_{j}^4\,,
\end{equation}
where we have neglected terms of order $(\epsilon\pi_{0}^2)^{-1}$ with respect to order $1$ terms and where we have used equation \eqref{eqn:phaseequation}.
Of course, the second term on the right-hand-side above, once inserted in equation \eqref{eqn:deltamomentum}, would give a vanishing contribution when the condition $\sum_j Q_j=0$ is imposed. As discussed in the previous subsection, however, we will retain it in the computations below. Therefore, we have
\begin{align}
    &\delta \Pi^2_{\sigma_\epsilon;\chi_0,\pi_0}=  \frac{1}{4}\sum_{j}\mathcal{N}_{\epsilon}^2\int\diff\chi\frac{([\rho_j^2]')^2}{\rho_{j}^2}e^{-\frac{(\chi-\chi_{0})^2}{\epsilon}}\nonumber\\
&\qquad+\sum_{j}\mathcal{N}_{\epsilon}^2\int\diff\chi\rho_{j}^2e^{-\frac{(\chi-\chi_{0})^2}{\epsilon}}\left[\frac{1}{\epsilon}-\frac{(\chi-\chi_0)^2}{\epsilon^2}+\pi_0^2\right]\nonumber\\
&\qquad+\sum_{j}\mathcal{N}_{\epsilon}^2\int\diff\chi\frac{Q_j^2}{\rho_{j}^2}e^{-\frac{(\chi-\chi_{0})^2}{\epsilon}}+\sum_jQ_j\label{eqn:deltapifirstsimply} \, ,
\end{align}
where we have used that
\begin{align*}
    &\sum_{j}\mathcal{N}_{\epsilon}^2\int\diff\chi[\rho_j^2]'\frac{\chi-\chi_0}{\epsilon}e^{-\frac{(\chi-\chi_{0})^2}{\epsilon}}\\
    &\quad=-\sum_{j}\mathcal{N}_{\epsilon}^2\int\diff\chi\rho_j^2\left[\epsilon^{-1}-2\frac{(\chi-\chi_0)^2}{\epsilon^2} \right]e^{-\frac{(\chi-\chi_{0})^2}{\epsilon}}
\end{align*}
to simplify the cross-product term in the smaller square brackets in \eqref{eqn:deltamomentum}. Moreover, by using that
\begin{equation*}
    ([\rho_j^2]')^2=4\rho_j^2(\rho_j')^2=4\rho_j^2\left(\mathcal{E}_j-\frac{Q_j^2}{\rho_j^2}+\mu_j^2\rho_j^2\right)\,,
\end{equation*}
we notice that the term in the first line in \eqref{eqn:deltapifirstsimply} becomes 
\begin{equation*}
    \sum_j\mathcal{N}_{\epsilon}^2\int\diff\chi \left(\mathcal{E}_j-\frac{Q_j^2}{\rho_j^2}+\mu_j^2\rho_j^2\right)e^{-\frac{(\chi-\chi_{0})^2}{\epsilon}}\,,
\end{equation*}
so that we can actually rewrite \eqref{eqn:deltapifirstsimply} as
\begin{align}\label{eqn:secondsimplifydeltapi}
    \delta \Pi^2_{\sigma_\epsilon;\chi_0,\pi_0}&=\sum_{j}\mathcal{N}_{\epsilon}^2\int\diff\chi\rho_{j}^2e^{-\frac{(\chi-\chi_{0})^2}{\epsilon}}\left[\frac{1}{\epsilon}+\mu_j^2+\pi_0^2\right]\nonumber\\&\quad-\sum_{j}\mathcal{N}_{\epsilon}^2\int\diff\chi\rho_{j}^2\frac{(\chi-\chi_0)^2}{\epsilon^2}e^{-\frac{(\chi-\chi_{0})^2}{\epsilon}}\nonumber\\&\quad+\sum_j\left(\mathcal{E}_j+Q_j\right)\,.
\end{align}
Let us now evaluate the term in the second line above. By the standard expansion of $\rho_j^2$ in powers of $(\chi-\chi_0)$ and the usual Gaussian integrations, we have that it is equal to
\begin{equation*}
    -\sum_j\frac{\rho_j^2(\chi_0)}{2\epsilon}\left(1+\sum_{n=1}^\infty\frac{[\rho_j^2]^{(2n)}(\chi_0)}{\rho_j^2(\chi_0)}\frac{\epsilon^n}{n!4^n}(2n+1)\right)\,.
\end{equation*}
On the other hand, the first term of equation \eqref{eqn:secondsimplifydeltapi} can be computed in the very same way as we did for the number operator, giving
\begin{equation*}
    \sum_j\left[\frac{1}{\epsilon}+\mu_j^2+\pi_0^2\right]\rho_j^2(\chi_0)\left(1+\sum_{n=1}^\infty\frac{[\rho_j^2]^{(2n)}(\chi_0)}{\rho_j^2(\chi_0)}\frac{\epsilon^n}{n!4^n}\right)\,,
\end{equation*}
which, combined with the equation above and inserted in \eqref{eqn:secondsimplifydeltapi} gives an expression for the relative variance, upon dividing the result by \eqref{eqn:averagepi}:
\begin{align*}
    &\delta^2_\Pi=\frac{1}{(\pi_0N+\sum_jQ_j)^2}\Biggl\{\sum_j\left(Q_j+\mathcal{E}_j\right)\\
    &\quad+\sum_j\left[\frac{1}{\epsilon}+\mu_j^2+\pi_0^2\right]\rho_j^2(\chi_0)\left[1+\sum_{n=1}^\infty\frac{[\rho_j^2]^{(2n)}(\chi_0)}{\rho_j^2(\chi_0)}\frac{\epsilon^n}{n!4^n}\right] \\
    &\quad-\sum_j\frac{\rho_j^2(\chi_0)}{2\epsilon}\left(1+\sum_{n=1}^\infty\frac{[\rho_j^2]^{(2n)}(\chi_0)}{\rho_j^2(\chi_0)}\frac{\epsilon^n}{n!4^n}(2n+1)\right)\Biggr\} \, .
\end{align*}

\paragraph*{Massless scalar field operator.}
Last, let us discuss the variance of the massless scalar field operator. First, we compute
\begin{align*}
    \delta X^2_{\sigma_\epsilon;\chi0,\pi_0}&\equiv \braket{\hat{X}^2}_{\sigma_\epsilon;\chi_0,\pi_0}-\braket{\hat{X}}_{\sigma_\epsilon;\chi_0,\pi_0}\\&=\sum_j\int\diff\chi\,\chi^2\rho_j^2(\chi)\vert\eta_\epsilon(\chi-\chi_0;\pi_0)\vert^2\,.
\end{align*}
As usual, we can expand the right-hand-side around $\chi_0$ in order to evaluate this quantity. We obtain
\begin{widetext}
\begin{align}
   \delta X^2_{\sigma_\epsilon;\chi_0,\pi_0}&=\chi_0^2\sum_j\rho_j^2(\chi_0)\int\diff\chi\,\vert\eta_\epsilon(\chi-\chi_0;\pi_0)\vert^2\nonumber\\
   &\quad\times\left\{1+\sum_{n=1}^\infty\left[\frac{[\rho_j^2]^{(2n)}(\chi_0)}{\rho_j^2(\chi_0)}+4n \frac{[\rho_j^2]^{(2n-1)}(\chi_0)}{\chi_0\rho_j^2(\chi_0)}+2n(2n-1)\frac{[\rho_j^2]^{(2n-2)}(\chi_0)}{\chi_0^2\rho_j^2(\chi_0)}\right]\frac{(\chi-\chi_0)^{2n}}{(2n)!}\right\}\nonumber\\
   &=\chi_0^2\sum_j\rho_j^2(\chi_0)\left\{1+\sum_{n=1}^\infty\left[\frac{[\rho_j^2]^{(2n)}(\chi_0)}{\rho_j^2(\chi_0)}+4n \frac{[\rho_j^2]^{(2n-1)}(\chi_0)}{\chi_0\rho_j^2(\chi_0)}+2n(2n-1)\frac{[\rho_j^2]^{(2n-2)}(\chi_0)}{\chi_0^2\rho_j^2(\chi_0)}\right]\frac{\epsilon^n}{4^n n!}\right\} \, .
\end{align}
\end{widetext}
The relative variance of $\hat{\chi}$ is then obtained by dividing this quantity by equation \eqref{eqn:completemassless} squared.
\section{Classical relational dynamics of flat FRW spacetime}\label{app:classical}
Following \cite{Marchetti:2020umh}, we review here the classical relational setting, with the purpose of comparing the effective relational results described above with the classical ones. The starting point is the total Hamiltonian of a flat FRW spacetime with a minimally coupled massless scalar field \cite{Bojowald:2008zzb}:
\begin{align*}
S&=\frac{3}{8\pi G}\int\diff t\,N\left(-\frac{aV_{0}\dot{a}^2}{N^2}+\frac{V}{N}\frac{\dot{\chi}^2}{2N}\right)\\
&=-\frac{3}{8\pi G}\int\diff t NV\left(\frac{H^2}{N^2}-\frac{4\pi G}{3}\frac{\dot{\chi^2}}{N^2}\right)\,.
\end{align*}
Here $\chi$ represents the massless scalar field, a dot denotes a derivative with respect to cosmic time $t$ and $V_{0}$ is the fiducial coordinate volume (so that $V\equiv V_{0}a^3$). Performing an Hamiltonitn analysis of the above action, one obtains the following constraint:
\begin{equation}
\mathcal{C}=-\frac{3}{8\pi G}NVH^2+\frac{N\pi_{\chi}^2}{2V}=0\,,
\end{equation}
which, with the Poisson brackets $\{H,V\}=4\pi G$ and $\{\chi,\pi_{\chi}\}=1$, implies that the equation of motion for the massless scalar field and the volume are
\begin{equation*}
\dot{\chi}=\{\chi,\mathcal{C}\}=N\pi_\chi/V\,,\qquad
\dot{V}=\{V,\mathcal{C}\}=3NVH\,.
\end{equation*}
Using the first one into the second one, together with the constraint equation, one gets
\begin{subequations}\label{eqn:friedmannrelational}
\begin{equation}
\left(\frac{1}{3V}\frac{\diff V}{\diff\chi}\right)^2\equiv\left(\frac{V'}{3V}\right)^2=\frac{4\pi G}{3}\,.
\end{equation}
This equation can be then derived with respect to $\chi$ to obtain we find
\begin{equation}
V''/V=\left(V'/V\right)^2=12\pi G\,.
\end{equation}
\end{subequations}
These are the relational equations for a spatially flat FRW spacetime.
\paragraph*{Gauge fixing.}
The same kind of dynamics can be obtained through a gauge fixing, choosing $\chi$ as our time variable, i.e.,  fixing $N=V\dot{\chi}/\pi_{\chi}$:
\begin{align}
S&=-\frac{3}{8\pi G}\int\diff t\dot{\chi} \frac{V^2}{\pi_{\chi}}\left(\frac{H^2\pi_{\chi}^2}{V^2\dot{\chi}^2}-\frac{4\pi G}{3}\frac{\dot{\chi^2}\pi_{\chi}^2}{V^2\dot{\chi}^2}\right)\nonumber\\
&=-\frac{3\pi_{\chi}}{8\pi G}\int\diff\chi\left(\mathcal{H}^2-\frac{4\pi G}{3}\right)\,.
\end{align}
Using that $\mathcal{H}=V'/(3V)$ one obtains the following equations of motion:
 \begin{equation*}
  V''/V=(V')^2/V^2\,,
  \end{equation*}
which is the second Friedmann equation, and which gives indeed the same dynamics as before. Neglecting unimportant constant terms, the Hamiltonian obtained from the above Lagrangian can be written as
\begin{equation}\label{eqn:classicalrelationalhamiltonian}
H_{\text{rel}}=-\frac{3\pi_{\chi}}{8\pi G}\mathcal{H}^2\,.
\end{equation}

\bibliographystyle{jhep}
\bibliography{fluctuations}

\end{document}